\documentclass[aps,prd,onecolumn,notitlepage,nofootinbib,superscriptaddress]{revtex4-2}

\usepackage[T1]{fontenc}
\usepackage{amsmath,amssymb,mathtools,bm}
\usepackage{graphicx}
\usepackage{booktabs}
\usepackage{array}
\usepackage{siunitx}
\usepackage{xcolor}
\usepackage{hyperref}
\usepackage{subcaption}

\graphicspath{{figures/}}
\hypersetup{
  colorlinks=true,
  citecolor=blue,
  linkcolor=blue,
  urlcolor=blue
}

\newcommand{\GeV}{\,\mathrm{GeV}}
\newcommand{\TeV}{\,\mathrm{TeV}}
\newcommand{\keV}{\,\mathrm{keV}}
\newcommand{\eV}{\,\mathrm{eV}}

\newcommand{\Lag}{\mathcal L}

\begin{document}

\title{Modular $S_4$ Inverse Seesaw at the LZ High-Recoil Frontier}

\author{Waqas Ahmed}
\email{waqasmit@hbpu.edu.cn}
\affiliation{Center for Fundamental Physics and School of Artificial Intelligence,
Hubei Polytechnic University, Huangshi 435003, China}
\author{Vasileios Basiouris}
\email{v.basiouris@gmail.com}
\author{George K. Leontaris}
\email{leonta@uoi.gr}

\affiliation{Department of Physics, University of Ioannina,
45110 Ioannina, Greece}

\begin{abstract}
The extension of the LUX--ZEPLIN nuclear-recoil analysis to
approximately $270\keV$ probes a qualitatively new dark-matter regime in
which endothermic scattering can suppress conventional low-energy recoils
and concentrate the surviving spectrum near the Galactic high-velocity
tail.  We construct a modular-$S_4$ realization in which a minimal
$2N_R+2S$ inverse seesaw accounts for lepton flavor, while a separate pair
of Standard-Model-singlet fermions forms the pseudo-Dirac state relevant
for high-recoil xenon scattering.  The sectors share a common complex
modulus $\tau$, so the modular forms selected by charged-lepton and
neutrino data also enter the operators generating the dark Dirac mass and
Majorana splitting. Our scan yields $181447$ lepton-sector solutions satisfying $\chi^2_{\ell+\nu}<30$, distributed among several disconnected modular regions, and predicts one effectively massless light neutrino at leading order.  Restoring the physical scales allows TeV-scale heavy
quasi-Dirac neutral fermions with small non-unitarity. We subsequently scan $2\times10^6$ dark-sector points only over flavor-accepted values of $\tau$ and identify common-modulus solutions in the LZ-oriented region
$300\GeV\le m_\chi\le2\TeV$ and $300\keV\le\delta\le400\keV$. An exploratory two-bin Poisson likelihood, including endothermic xenon
kinematics, the Helm form factor, Standard-Halo-Model velocities,
Gaussian energy smearing, and profiling over the signal normalization,
gives a best grid point at $(m_\chi,\delta)\simeq(1.847\TeV,371.5\keV)$,
close to the terrestrial inelastic kinematic ceiling.  The construction
provides a direct link between modular lepton flavor, TeV-scale
inverse-seesaw dynamics, and high-recoil inelastic dark matter.
\end{abstract}

\maketitle

\section{Introduction}

The LUX--ZEPLIN (LZ) Collaboration has recently extended its nuclear-recoil
analysis to energies of approximately $270\keV$ in an exposure of
$2.84$ tonne yr. One event is reconstructed at
$E_R=248\pm23_{\rm stat}\pm23_{\rm sys}\keV$, in a region with a small
known-background expectation. The collaboration reports a maximum local
significance of $3.4\sigma$ and a global significance of $2.6\sigma$ after
the look-elsewhere effect \cite{LZ2026HighRecoil}. The result is therefore
not evidence for dark matter by itself, but it opens a qualitatively new
direct-detection regime in which non-standard momentum dependence,
inelastic thresholds, nuclear-response effects, and the high-velocity tail
of the Galactic halo can become considerably more important than in the
conventional low-energy spin-independent search.

The event has consequently triggered a broad phenomenological response.
Endothermic interpretations are particularly well motivated because an
excitation energy of a few hundred keV can suppress the otherwise dominant
low-energy recoils while retaining support around the observed event
\cite{SuYangYang2026,DiMauro2026,PalmisanoTammaroTesi2026,
FanHeWangZhao2026,DelepineKhalil2026}. The proximity of the preferred
endothermic solutions to the Galactic speed boundary also makes the signal
sensitive to the high-velocity halo population and can lead to unusually
large annual modulation \cite{McCabe2026}. These observations emphasize
that the LZ high-recoil event probes not only the interaction strength but
also the relation among the dark-matter mass, the excitation energy, the
target nucleus, and the astrophysical velocity distribution.

A wide variety of ultraviolet and semi-ultraviolet realizations of this
kinematics have already been considered. Dark-photon constructions provide
one possibility \cite{Yamashita2026}, while electroweak multiplets
\cite{SmirnovGriffithBeacom2026}, singlet--doublet fermions
\cite{BorahSingletDoublet2026,LeeYoun2026}, baryon-flavor gauge symmetries
\cite{LeeBaryonFlavor2026}, radiative neutrino-mass models
\cite{OkadaShigekamiWu2026,BorahRadiativeDirac2026}, modular
inverse-seesaw/self-interacting dark matter
\cite{DasKarmakarMahapatraPaul2026}, hadrophilic freeze-in
\cite{HeHongJeesun2026}, composite dark matter
\cite{SanninoTurner2026}, and higher-dimensional operators
\cite{ChoudhuryDasSachdeva2026} provide alternative realizations.
Complementary explanations based on exothermic scattering
\cite{BaerBarger2026}, momentum-dependent elastic interactions
\cite{Unwin2026,KhanEtAl2026,CentellesChuliaEtAl2026}, and dark-matter
absorption \cite{LouLu2026} have also been proposed.

Supersymmetric and extra-dimensional constructions are particularly useful
for addressing the origin of the small mass splitting rather than treating
it as an independent low-energy parameter. In GmSUGRA, a thermal Higgsino
branch can dynamically generate sub-MeV neutralino splittings extending
through the LZ-motivated region; the corresponding few-hundred-keV branch
is simultaneously subject to solar-neutrino constraints, while the same
high-scale construction also admits larger splittings
\cite{KhanAhmedLiRazaMuhammad2026}. Extra-dimensional approaches provide
a different mechanism. A five-dimensional Dark-Dimension construction can
generate the small Majorana splitting through geometric suppression and
leads to characteristic multi-target correlations
\cite{AhmedLeontaris2026}. More recently, a localized pseudo-Dirac singlet
coupled to a bulk $U(1)_D$ field was shown to correlate the mass splitting,
fermion localization, and the full Kaluza--Klein dark-photon mediator
tower, thereby modifying the high-recoil xenon spectrum relative to a
single-mediator description \cite{AhmedLeontarisKK2026}.

There is also an important nuclear aspect to the kinematic-edge regime.
When $v_{\min}$ approaches the truncation of the Galactic velocity
distribution, the small differences among the physical xenon isotope
masses are magnified. An isotope-resolved treatment shows that the relative
contributions of the xenon isotopes can therefore differ substantially from
their natural abundances and can be further reorganized by the Helm form
factor at large momentum transfer \cite{AhmedAhmadRehman2026}. This
``isotope filtering'' effect provides an additional reason not to regard
the high-recoil event as merely an extrapolation of the standard
low-energy xenon analysis.

\textcolor{black}{
The same LZ-motivated parameter space is also being tested through
dark-matter capture in the Sun and the associated high-energy neutrino flux.
For a nearly pure thermal Higgsino close to the conventional
$m_\chi\simeq1.1\TeV$ freeze-out mass, several analyses find strong
solar-neutrino constraints on the few-hundred-keV inelastic splitting
relevant to the LZ event
\cite{PospelovRamani2026,BoseEtAlHiggsino2026}. These results, however,
do not constitute a model-independent exclusion of the Higgsino
interpretation. The capture efficiency depends on the Higgsino mass,
the inelastic splitting, the high-velocity halo distribution, elastic
cooling processes inside the Sun, the annihilation channels, and the
treatment of the LZ high-energy sideband.}

\textcolor{black}{In particular, dedicated heavy-Higgsino studies show that regions
compatible with both the LZ high-recoil candidate and IceCube can reappear
once the Higgsino mass is increased
\cite{Langhoff2026,FreeseTheodosopoulos2026}. Depending on the halo and
solar assumptions, the surviving parameter space can extend from masses of
order a few TeV to much larger values while retaining inelastic splittings
of a few hundred keV. Thus the solar-neutrino constraint should be viewed
as a discriminator among different Higgsino masses and thermal histories,
rather than as a universal exclusion of the Higgsino interpretation.}

\textcolor{black}{A complementary possibility is provided by a nonstandard cosmological
history. Ref.~\cite{Pallis2026HeavyHiggsino} showed that a heavy Higgsino
with $m_{\widetilde h}\sim0.1\,{\rm PeV}$ can be compatible with the
LZ high-recoil interpretation while reproducing the observed cold-dark-matter
abundance for a sufficiently low reheating temperature,
$T_{\rm rh}\lesssim595\,{\rm GeV}$. In this realization the final abundance
receives an important nonthermal contribution from the out-of-equilibrium
decay of a heavier modulus field. This provides an explicit example in
which the relic-density requirement does not force the LZ interpretation
back to the conventional $m_\chi\simeq1.1\TeV$ thermal-Higgsino regime.
More generally, these results emphasize that solar-neutrino constraints
select particular combinations of mass, splitting, interaction structure,
halo assumptions, and cosmological history rather than imposing a
universal upper bound on inelastic dark matter.
}

The mechanism studied in the present work is the endothermic transition
\begin{equation}
  \chi_1+N\longrightarrow\chi_2+N,
  \qquad
  \delta\equiv m_{\chi_2}-m_{\chi_1}>0.
  \label{eq:endothermic}
\end{equation}
Inelastic dark matter was originally introduced in a different
direct-detection context
\cite{TuckerSmithWeiner2001,TuckerSmithWeiner2005}, and its potential
importance at high recoil energies was emphasized in
Ref.~\cite{BramanteEtAl2016}. For $\delta>0$, part of the incoming kinetic
energy is used to excite the heavier state. Consequently the minimum
incident speed is raised, the low-energy spectrum is strongly suppressed,
and the surviving events are driven toward the edge of the Galactic
velocity distribution. This is precisely the qualitative behavior favored
by the LZ high-recoil candidate.

Our objective is to go beyond a purely kinematic description and investigate
whether the TeV-scale pseudo-Dirac state and its few-hundred-keV splitting
can be embedded into a predictive flavor framework. Modular symmetries are
well suited to this purpose because Yukawa structures are promoted to
holomorphic functions of a complex modulus $\tau$, substantially reducing
the need for large flavon sectors and allowing seemingly unrelated
observables to be correlated through a common modulus
\cite{Feruglio2017,PenedoPetcov2019,NovichkovEtAl2019}. At level four,
the finite modular group is
\begin{equation}
\Gamma_4\simeq S_4,
\end{equation}
which provides a particularly economical structure for lepton masses and
mixing.

Inverse-seesaw models provide the second ingredient of our construction.
In contrast to the conventional high-scale seesaw, the inverse seesaw
allows the heavy neutral fermions to remain at the electroweak or TeV scale
because the smallness of the active-neutrino masses is controlled by a
separate, technically natural lepton-number-violating parameter
\cite{MohapatraValle1986,BernabeuEtAl1987,MalinskyEtAl2009}. A modular
$S_4$ inverse seesaw has previously been studied in connection with lepton
mixing and a separate keV sterile state as warm dark matter
\cite{ZhangZhou2021}. Such a keV state, however, cannot generate a
$\sim250\keV$ xenon nuclear recoil through ordinary nonrelativistic
scattering. The neutrino and high-recoil dark sectors must therefore be
distinguished.

We consequently adopt a minimal $2N_R+2S$ inverse-seesaw sector to account
for neutrino masses and mixing, while introducing a separate pair of
Standard-Model-singlet fermions $\chi_{1,2}$ for the TeV-scale
pseudo-Dirac dark matter. The two sets of neutral fermions are physically
distinct and their small Majorana parameters are not identified. Their
connection instead arises from the common modular background: the same
complex modulus $\tau$ that determines the charged-lepton and
inverse-seesaw textures also enters the higher-dimensional operators that
generate the dark Dirac mass and its small Majorana splitting.

The central logic of the construction is therefore
\begin{equation}
\boxed{
\begin{aligned}
S_4\ {\rm modular\ flavor}
&\longrightarrow {\rm inverse\ seesaw}
\longrightarrow \tau_{\rm flavor}
\\[-1mm]
&\longrightarrow {\rm pseudo\!-\!Dirac\ dark\ sector}
\longrightarrow {\rm LZ\ high\ recoil}.
\end{aligned}}
\label{eq:logic}
\end{equation}
The common-modulus requirement is the predictive bridge between the two
sectors: the dark-sector scan is carried out only for values of $\tau$
already accepted by the charged-lepton and neutrino fit. The resulting
dark-matter spectrum is therefore filtered by the flavor structure rather
than generated by an independent scan over the full modular domain.

The paper is organized as follows. We first introduce the modular $S_4$
flavor framework and construct the charged-lepton sector. The inverse
seesaw is then discussed separately, including the full $7\times7$ neutral
mass matrix, the rank-two light-neutrino structure, heavy quasi-Dirac
states, and non-unitarity. We subsequently introduce the pseudo-Dirac
singlet dark sector and derive the relation between the low-energy mass
splitting and the ultraviolet scales. We then present the global--local
flavor scan, the physical inverse-seesaw parameter space, and the
common-modulus dark solutions. Finally, we analyze the xenon endothermic
kinematics, recoil spectra, detector smearing, and an exploratory LZ
two-bin likelihood. Technical derivations are collected in the appendices.

\section{Inverse-seesaw neutrino sector}

{\color{black}
We now construct the neutral-fermion sector associated with the
$2N_R+2S$ field content summarized in Table~\ref{tab:visible}.  In
addition to the three active neutrinos, the model contains two
right-handed neutrinos,
$N_R=(N_R^1,N_R^2)$, and two sterile singlets,
$S=(S_1,S_2)$.  This realizes the minimal inverse-seesaw structure with
two heavy quasi-Dirac pairs.  The inverse-seesaw mechanism was introduced
in Refs.~\cite{MohapatraValle1986,BernabeuEtAl1987}, while TeV-scale
realizations and the associated non-unitarity effects have been studied
extensively
\cite{MalinskyEtAl2009,FernandezMartinezEtAl2016,BlennowEtAl2023}.
Modular inverse-seesaw constructions based on $S_4$ have also been
considered previously, including scenarios with an additional sterile
state playing the role of warm dark matter
\cite{ZhangZhou2021}.  In the present model, however, the minimal
$2N_R+2S$ sector is used only for neutrino-mass generation and the
corresponding heavy quasi-Dirac spectrum.
{\renewcommand{\arraystretch}{1.1}
\begin{table}[h] 
\makebox[\textwidth][c]{
		\begin{tabular}{ccccc}
			\hline\hline
			Fields  & $S_4$ & $-k$ \\
            \hline 
            $L_{1,2},e^c_{1,2} $  & $2$ & $(4,2)$\\
            $L_3,e^c_{3} $  & $1$ & $(6,2)$\\
            $H_u$  & $1$ & $0$\\
            $H_d$  & $1$ & $0$\\
            $N_R$  & $2$ & $0$\\
            $S$    & $2$ & $4$\\
			\hline\hline
		\end{tabular}
      }
  \caption{\small \textcolor{black}{Modular weights, and $S_4$ discrete modular group representations associated with the chiral superfields hosted in the model~. If we are not interested in the leptonic sector, then we do not need the $e^c,H_d$ fields~.}}
  \label{tab:visible}
\end{table}}
\noindent \textcolor{black}{Starting from the leptonic sector, the superpotential in this sector can be written as}
\begin{align}
    W_l = \alpha (e^c_{1,2} L_{1,2} Y^{(6)}_{(1,1',2)})_1 H_d + \beta ( e^c_3 L_3 Y_{1}^{(8)})_1 H_d~,
\end{align}
\textcolor{black}{where if we expand in the components of the modular forms, the relevant couplings are given by}
\begin{align}\label{eq:Yell}
    W_l = &\alpha( e_1^c L_1 + e^c_2 L_2)Y_1^{(6)} + \alpha_1(e_1^c L_2 - e^c_2 L_1)Y_{1'}^{(6)} + \alpha_2S_Y \big[ (e_2^c L_2 - e_1^c L_1)Y_1 + ( e_2^c L_1 + e_1^c L_2)Y_2 \big] + \notag \\
    & + \beta e^c_3 L_3 Y_1^{(8)}~.
\end{align}
The modular-invariant superpotential relevant for the neutral sector can
be written as
\begin{align}
W_\nu={}&
c_1(N_RL_D)_{\mathbf1}Y_{\mathbf1}^{(4)}H_u
+c_2[(N_RL_D)_{\mathbf2}Y_{\mathbf2}^{(4)}]_{\mathbf1}H_u
\nonumber\\
&+
c_3S_Y(N_R^1Y_1+N_R^2Y_2)L_3H_u
+d_1S_Y(N_R^1S_1+N_R^2S_2)
\nonumber\\
&+
d_2\left[
D_Y(N_R^2S_2-N_R^1S_1)
+P_Y(N_R^2S_1+N_R^1S_2)
\right]
+W_\mu ,
\label{eq:Wnu}
\end{align}
where the small lepton-number-violating contribution is
\begin{align}
W_\mu={}&
\frac12 e_1(S_1S_1+S_2S_2)S_Y^2
\nonumber\\
&+
\frac12 e_2S_Y
\left[
(S_2S_2-S_1S_1)D_Y
+
(S_2S_1+S_1S_2)P_Y
\right].
\label{eq:Wmu}
\end{align}
The second term in Eq.~\eqref{eq:Wmu} originates from the unique
doublet contraction
\begin{equation}
\frac12 e_2
\left[
(SS)_{\mathbf2}Y_{\mathbf2}^{(8)}
\right]_{\mathbf1},
\end{equation}
where
\begin{equation}
Y_{\mathbf2}^{(8)}
=
S_Y
\begin{pmatrix}
D_Y\\
P_Y
\end{pmatrix}.
\label{eq:Y28}
\end{equation}
Hence the structures proportional to $D_YS_Y$ and $P_YS_Y$ are the two
components of the same modular doublet and are governed by the common
coefficient $e_2$. \textcolor{black}{The structure of the modular forms defined above can be found in the following work \cite{Wang:2020dbp,Basiouris:2024qfe,Okada:2026fef} and in Appendix~\ref{app:modforms}.}

After electroweak symmetry breaking, the Dirac mass matrix connecting the
active and right-handed neutrinos is
\begin{equation}
M_D=v_u
\begin{pmatrix}
c_1S_Y+c_2D_Y & c_2P_Y\\
c_2P_Y & c_1S_Y-c_2D_Y\\
c_3S_YY_1 & c_3S_YY_2
\end{pmatrix},
\label{eq:MD}
\end{equation}
while the lepton-number-conserving mass matrix connecting $N_R$ and $S$
takes the form
\begin{equation}
M_R=
\begin{pmatrix}
d_1S_Y-d_2D_Y & d_2P_Y\\
d_2P_Y & d_1S_Y+d_2D_Y
\end{pmatrix}.
\label{eq:MR}
\end{equation}
The sterile Majorana block is
\begin{equation}
\mu_S=
\begin{pmatrix}
e_1S_Y^2-e_2D_YS_Y &
e_2P_YS_Y\\
e_2P_YS_Y &
e_1S_Y^2+e_2D_YS_Y
\end{pmatrix}.
\label{eq:muS}
\end{equation}

The coefficients $d_i$ and $e_i$ are dimensionful effective parameters.
The former determine the characteristic lepton-number-conserving
inverse-seesaw scale, whereas the latter control the much smaller
lepton-number-violating scale.  This treatment is consistent with the
modular construction because the corresponding fermion bilinears are
accompanied by modular forms carrying the appropriate modular weights.
Moreover, the hierarchy between these two scales is technically natural:
in the limit $\mu_S\rightarrow0$, the approximate lepton-number symmetry
is restored and the heavy neutral fermions combine into exact Dirac pairs.

At the effective level, we characterize the two relevant scales by
$M_0$ and $\mu_0$.  They may alternatively arise from vacuum expectation
values of additional modular-singlet fields,
\begin{equation}
M_0=\lambda_R\langle X_R\rangle,
\qquad
\mu_0=\lambda_\mu\langle X_\mu\rangle,
\label{eq:vevcompletion}
\end{equation}
providing a possible ultraviolet origin of the inverse-seesaw masses.
Since the detailed scalar potential responsible for such symmetry
breaking does not affect the low-energy neutrino structure considered
here, we treat $M_0$ and $\mu_0$ as effective mass parameters.}

In the basis $(\nu_L,N_R^c,S)$, the full neutral-fermion mass matrix is
\begin{equation}
\mathcal M_\nu=
\begin{pmatrix}
0_{3\times3} & M_D & 0_{3\times2}\\
M_D^T & 0_{2\times2} & M_R^T\\
0_{2\times3} & M_R & \mu_S
\end{pmatrix}.
\label{eq:M7}
\end{equation}
The vanishing $N_RN_R$ block follows from the approximate lepton-number
symmetry, whereas $\mu_S$ provides the small $\Delta L=2$ breaking that
characterizes the inverse seesaw.

For
\begin{equation}
\|\mu_S\|\ll\|M_R\|,
\end{equation}
block diagonalization yields the effective light-neutrino mass matrix
\begin{equation}
m_\nu\simeq
M_DM_R^{-1}\mu_SM_R^{-T}M_D^T.
\label{eq:ISS}
\end{equation}
A derivation is given in Appendix~\ref{app:iss}.  Since $M_D$ is a
$3\times2$ matrix,
\begin{equation}
{\rm rank}(m_\nu)\leq2,
\label{eq:rank}
\end{equation}
so that one light neutrino is massless at leading order.  This prediction
follows directly from the minimal $2N_R+2S$ particle content and does not
depend on the detailed values of the model parameters.

To disentangle the flavor structure from the overall mass scales, we write
\begin{equation}
c_1=c_0,\qquad
c_2=c_0r_{c2},\qquad
c_3=c_0r_{c3},
\end{equation}
together with
\begin{equation}
d_1=M_0,\qquad
d_2=M_0r_d,
\end{equation}
and
\begin{equation}
e_1=\mu_0,\qquad
e_2=\mu_0r_{e2}.
\end{equation}
The three mass matrices then factorize as
\begin{equation}
M_D=v_uc_0\widehat Y_D,\qquad
M_R=M_0\widehat Y_R,\qquad
\mu_S=\mu_0\widehat Y_\mu,
\end{equation}
where the hatted matrices depend only on the modulus $\tau$ and on
dimensionless coupling ratios.  The light-neutrino mass matrix can
therefore be written as
\begin{equation}
m_\nu=
\frac{v_u^2c_0^2\mu_0}{M_0^2}\,
\widehat m_\nu
\equiv
\kappa_\nu\,\widehat m_\nu,
\label{eq:kappanu}
\end{equation}
with
\begin{equation}
\widehat m_\nu=
\widehat Y_D\widehat Y_R^{-1}
\widehat Y_\mu
\widehat Y_R^{-T}\widehat Y_D^T.
\end{equation}
Thus the overall neutrino-mass scale is controlled by
\begin{equation}
\kappa_\nu=
\frac{v_u^2c_0^2\mu_0}{M_0^2},
\end{equation}
whereas the flavor structure is determined by
\begin{equation}
\left\{
\tau,\,
r_{\ell2},r_{\ell3},\,
r_{c2},r_{c3},r_d,r_{e2}
\right\}.
\label{eq:shapepars}
\end{equation}
The common modulus $\tau$ therefore enters both the charged-lepton and
neutrino sectors, while $c_0$, $M_0$, and $\mu_0$ set the corresponding
physical mass scales. The neutrino mixing matrix is obtained from
\begin{equation}
U_\nu^\dagger
(m_\nu m_\nu^\dagger)
U_\nu
=
{\rm diag}(m_1^2,m_2^2,m_3^2),
\end{equation}
and the physical leptonic mixing matrix is
\begin{equation}
U_{\rm PMNS}=U_\ell^\dagger U_\nu.
\end{equation}
The standard mixing observables are
\begin{align}
s_{13}^2&=|U_{e3}|^2,
\nonumber\\
s_{12}^2&=
\frac{|U_{e2}|^2}{1-|U_{e3}|^2},
\nonumber\\
s_{23}^2&=
\frac{|U_{\mu3}|^2}{1-|U_{e3}|^2},
\nonumber\\
R_\nu&=
\frac{\Delta m_{21}^2}{\Delta m_{31}^2}.
\label{eq:mixobs}
\end{align}
The Dirac CP phase is extracted from the Jarlskog invariant.  For the
phenomenological analysis we use
\begin{equation}
s_{12}^2=0.307,\qquad
s_{23}^2=0.572,\qquad
s_{13}^2=0.02203,
\end{equation}
and
\begin{equation}
R_\nu=0.029510,\qquad
\delta_{\rm CP}=197^\circ,
\label{eq:inputs}
\end{equation}
with the corresponding uncertainties specified in the numerical analysis.
Current global neutrino-oscillation results are summarized in
Ref.~\cite{NuFIT60}.

The leading active--heavy mixing matrix is
\begin{equation}
F=M_DM_R^{-1}
=
\frac{v_uc_0}{M_0}
\widehat Y_D\widehat Y_R^{-1},
\end{equation}
which induces a deviation from unitarity in the effective light-neutrino
mixing matrix.  At leading order,
\begin{equation}
\eta_{\rm NU}\simeq
\frac12FF^\dagger.
\label{eq:eta}
\end{equation}
We characterize the non-unitarity constraints through
\begin{equation}
R_\eta\equiv
\max_{ij}
\frac{|\eta_{ij}|}{\eta_{ij}^{\rm lim}},
\label{eq:Reta}
\end{equation}
and require
\begin{equation}
R_\eta<1.
\end{equation}
Current global constraints on non-unitarity and active--heavy neutrino
mixing can be found in
Refs.~\cite{FernandezMartinezEtAl2016,BlennowEtAl2023}.

\section{Pseudo-Dirac singlet dark sector}

The LZ dark state is not identified with an inverse-seesaw sterile
fermion.  Instead, we introduce two additional singlet superfields
$\chi_1,\chi_2$ and a symmetry-breaking field $\Phi$.  This separation
avoids a basic kinematic mismatch: keV sterile-neutrino dark matter
\cite{AbadaArcadiLucente2014,ZhangZhou2021} cannot generate a
$\sim250\keV$ xenon recoil through ordinary nonrelativistic scattering.

A representative assignment is
\begin{table}[b]
\centering
\caption{Dark-sector assignments.  The visible fields are neutral under
$U(1)_X$ and even under the remnant parity.}
\label{tab:dark}
\begin{tabular}{lcccc}
\toprule
field & $S_4$ & $-k$ & $U(1)_X$ & $Z_2$\\
\midrule
$\chi_1$ & $\mathbf1$ & $1$ & $+1$ & $-$\\
$\chi_2$ & $\mathbf1$ & $3$ & $+3$ & $-$\\
$\Phi$ & $\mathbf1$ & $0$ & $-2$ & $+$\\
Visible sector & Table I & -- & $0$ & $+$\\
\bottomrule
\end{tabular}
\end{table}
If $\Phi$ is the only field whose vacuum expectation value breaks $U(1)_X$
and its charge is even, $\exp(i\pi Q_X)$ remains unbroken.  The odd charges
of $\chi_{1,2}$ therefore realize a remnant $Z_2$, providing a symmetry
origin for dark-matter stability.  The use of a high-scale symmetry to
organize hierarchical operators is closely related in spirit to the
Froggatt--Nielsen mechanism \cite{Froggatt:1978nt}.

The leading operators are
\begin{equation}
W_{\rm DM}=
y_D\frac{\Phi^2}{\Lambda}
Y_{\mathbf1}^{(4)}(\tau)\chi_1\chi_2
+
\frac{y_M}{2}\frac{\Phi^3}{\Lambda^2}
Y_{\mathbf1}^{(6)}(\tau)\chi_2\chi_2.
\label{eq:WDM}
\end{equation}
For the adopted modular weights, a corresponding $11$ Majorana term is
absent at this order because no weight-2 $S_4$ singlet is available.
After $\langle\Phi\rangle=v_\Phi$,
\begin{equation}
M_\chi=
y_DY_{\mathbf1}^{(4)}(\tau)\frac{v_\Phi^2}{\Lambda},
\qquad
\mu_\chi=
y_MY_{\mathbf1}^{(6)}(\tau)\frac{v_\Phi^3}{\Lambda^2},
\end{equation}
and
\begin{equation}
\mathcal M_\chi=
\begin{pmatrix}
0&M_\chi\\
M_\chi&\mu_\chi
\end{pmatrix}.
\label{eq:Mchi}
\end{equation}
For $|\mu_\chi|\ll|M_\chi|$,
\begin{equation}
m_{\chi_{1,2}}\simeq
|M_\chi|\mp\frac{|\mu_\chi|}{2},
\qquad
m_\chi\simeq|M_\chi|,
\qquad
\delta\simeq|\mu_\chi|.
\label{eq:darksplit}
\end{equation}
The exact eigenvalues and the small-splitting expansion are derived in
Appendix~\ref{app:dark}.  The hierarchy is controlled by
\begin{equation}
\frac{\delta}{m_\chi}\simeq
\left|
\frac{y_MY_{\mathbf1}^{(6)}(\tau)}
     {y_DY_{\mathbf1}^{(4)}(\tau)}
\right|
\frac{v_\Phi}{\Lambda}.
\label{eq:ratio}
\end{equation}
Defining
$A=|y_DY_{\mathbf1}^{(4)}|$ and
$B=|y_MY_{\mathbf1}^{(6)}|$, the exact algebraic relations are
\begin{equation}
v_\Phi=
\frac{B}{A^2}\frac{m_\chi^2}{\delta},
\qquad
\Lambda=
\frac{B^2}{A^3}\frac{m_\chi^3}{\delta^2}.
\label{eq:UVexact}
\end{equation}
Thus a small inelastic splitting can be interpreted as a hierarchy of
physical scales modulated by calculable functions of the flavor modulus.

The high-scale $U(1)_X$ that organizes Eq.~\eqref{eq:WDM} need not be the
low-energy mediator of nuclear scattering.  A generic off-diagonal vector
portal can be written
\begin{equation}
\Lag_{\rm int}=
g_DV_\mu\overline{\chi_2}\gamma^\mu\chi_1
+
V_\mu\sum_qg_q\bar q\gamma^\mu q+\mathrm{H.c.}
\label{eq:portal}
\end{equation}
and a dark-photon realization can communicate with the visible sector
through kinetic mixing \textcolor{black}{$\epsilon$}
\begin{equation}
\Lag\supset-\frac{\epsilon}{2}B_{\mu\nu}V^{\mu\nu}
\label{eq:kinmix}
\end{equation}
\cite{Holdom:1985ag,FabbrichesiGabrielliLanfranchi2020}.  In the present work
the total direct-detection normalization is profiled in the likelihood,
so the primary predictions concern the spectral variables
$(m_\chi,\delta)$ and their modular origin rather than a unique mediator
benchmark.

\section{Flavor scan and physical inverse-seesaw parameter space}

We work in normal neutrino mass ordering throughout the numerical analysis. Every complex ratio in Eq.~\eqref{eq:shapepars} is parameterized as
\begin{equation}
r=10^\rho e^{i\phi},
\qquad
-2\le\rho\le\log_{10}5,
\qquad
-\pi\le\phi\le\pi.
\label{eq:rscan}
\end{equation}
We use $\tan\beta=5$, $v_{\rm EW}=174\GeV$, and hence
$v_u=170.621\GeV$.  The lepton fit is
\begin{align}
\chi^2_{\ell+\nu}={}&
\left(\frac{y_e^{\rm pred}-y_e}{0.03\,y_e}\right)^2
+\left(\frac{s_{12}^2-0.307}{0.012}\right)^2
+\left(\frac{s_{23}^2-0.572}{0.023}\right)^2
\nonumber\\
&+
\left(\frac{s_{13}^2-0.02203}{0.00058}\right)^2
+\left(\frac{R_\nu-0.029510}{0.000894}\right)^2
+\left(\frac{\Delta\delta_{\rm CP}}{41^\circ}\right)^2.
\label{eq:chi2}
\end{align}
We retain points with
\begin{equation}
\chi^2_{\ell+\nu}<30.
\label{eq:chi2cut}
\end{equation}

\begin{figure}[t]
\centering
\includegraphics[width=0.97\linewidth]{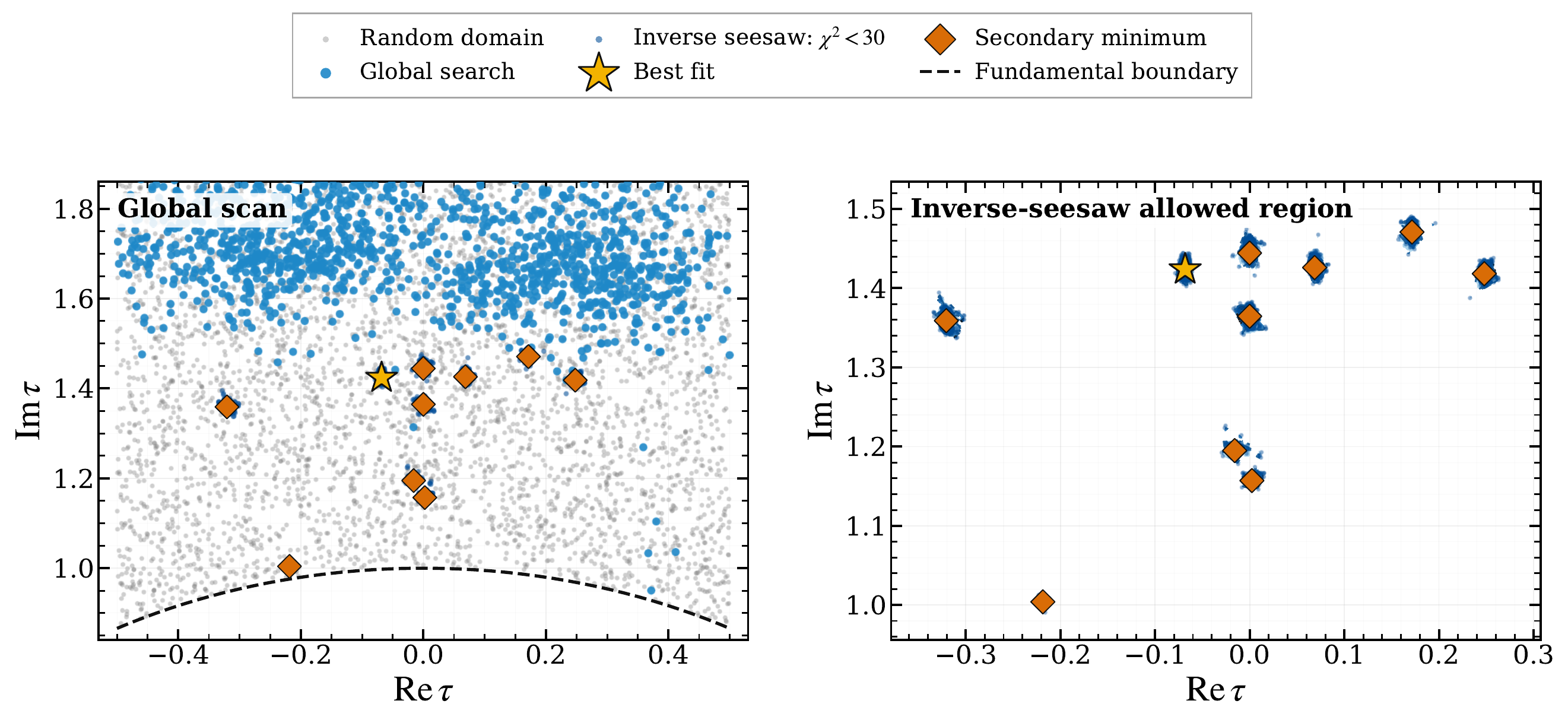}
\caption{Global and local modular parameter space.  The global search
explores the full fundamental region and finds several disconnected
low-$\chi^2$ basins.  The gold star marks the best fit and orange diamonds
mark secondary minima used to seed local scans.  The right panel displays
the dense inverse-seesaw islands satisfying $\chi^2_{\ell+\nu}<30$.
Filled regions are conservative scan envelopes of accepted points; they
should not be interpreted as a fixed-$\tau$ confidence construction.}
\label{fig:tau}
\end{figure}

In our scan, we perform 12 differential-evolution
runs over the full fundamental region, together with eight dedicated
negative-${\rm Re}\,\tau$ runs and eight dedicated
positive-${\rm Re}\,\tau$ runs.  Each run uses a population factor of 10
and 90 generations.  Up to ten distinct low-$\chi^2$ basins are retained
as local centers.  Around every center we generate 60000 fine, 35000 wide,
and 15000 broad proposals.  We then add 400000 adaptive core proposals
around already accepted points and 200000 edge proposals seeded from the
higher-$\chi^2$ part of the accepted sample.  Every adaptive proposal is
re-evaluated with the full charged-lepton plus neutrino objective and is
retained only when Eq.~\eqref{eq:chi2cut} is satisfied; interpolation is
used only to visualize the scan-supported envelopes in
Fig.~\ref{fig:tau}, not to create accepted solutions.

Figure~\ref{fig:tau} summarizes the modular parameter space.  The left
panel displays the full search geometry: gray points show the fundamental
domain, blue points show the final populations of the global optimizer,
and the dark accepted points satisfy Eq.~\eqref{eq:chi2cut}.  The global
population is not an exclusion map; it is a diagnostic of optimizer
coverage.  The right panel zooms into the densely sampled accepted basins.
The filled blue regions are scan-supported envelopes built from the actual
accepted points and masked outside regions supported by nearby samples.
They are therefore a visualization of the numerical solution islands, not
an exact fixed-$\tau$ profile likelihood.

Our scan contains
\begin{equation}
N_{\rm ISS}=181447
\end{equation}
accepted points with
\begin{equation}
-0.333893\lesssim{\rm Re}\,\tau\lesssim0.262356,
\qquad
0.990784\lesssim{\rm Im}\,\tau\lesssim1.489357.
\label{eq:taurange}
\end{equation}
Ten distinct local centers are explicitly refined.  Their locations and
fit qualities are summarized in Table~\ref{tab:centers}.
\begin{table}[t]
\centering
\caption{Distinct low-$\chi^2$ modular basins used as local centers in our scan.}
\label{tab:centers}
\begin{tabular}{ccc}
\toprule
${\rm Re}\,\tau$ & ${\rm Im}\,\tau$ & $\chi^2_{\ell+\nu}$\\
\midrule
$-0.068199$ & $1.423704$ & $4.87\times10^{-7}$\\
$+0.171487$ & $1.470816$ & $2\times10^{-6}$\\
$+0.068604$ & $1.425983$ & $1.3\times10^{-5}$\\
$-0.320524$ & $1.358962$ & $8.3\times10^{-5}$\\
$-0.000120$ & $1.444328$ & $1.32\times10^{-4}$\\
$+0.247844$ & $1.418259$ & $1.86\times10^{-4}$\\
$+0.002372$ & $1.157089$ & $8.41\times10^{-4}$\\
$-0.000002$ & $1.364570$ & $1.758\times10^{-3}$\\
$-0.015861$ & $1.194990$ & $8.947\times10^{-3}$\\
$-0.218497$ & $1.004092$ & $4.473\times10^{-2}$\\
\bottomrule
\end{tabular}
\end{table}
Both positive- and negative-${\rm Re}\,\tau$ branches are
explicitly refined and are included in the accepted sample.

At the best point,
\begin{equation}
s_{12}^2=0.30699938,\quad
s_{23}^2=0.57200226,\quad
s_{13}^2=0.02203000,
\end{equation}
\begin{equation}
R_\nu=0.02951017,\qquad
\delta_{\rm CP}=196.972^\circ,
\label{eq:bestobs}
\end{equation}
with
\begin{equation}
\tau_{\rm bf}=-0.0681990+1.4237036\,i,
\qquad
\chi^2_{\ell+\nu,\rm bf}=4.87\times10^{-7}.
\label{eq:besttau}
\end{equation}

\begin{figure}[t]
\centering
\includegraphics[width=0.48\linewidth]{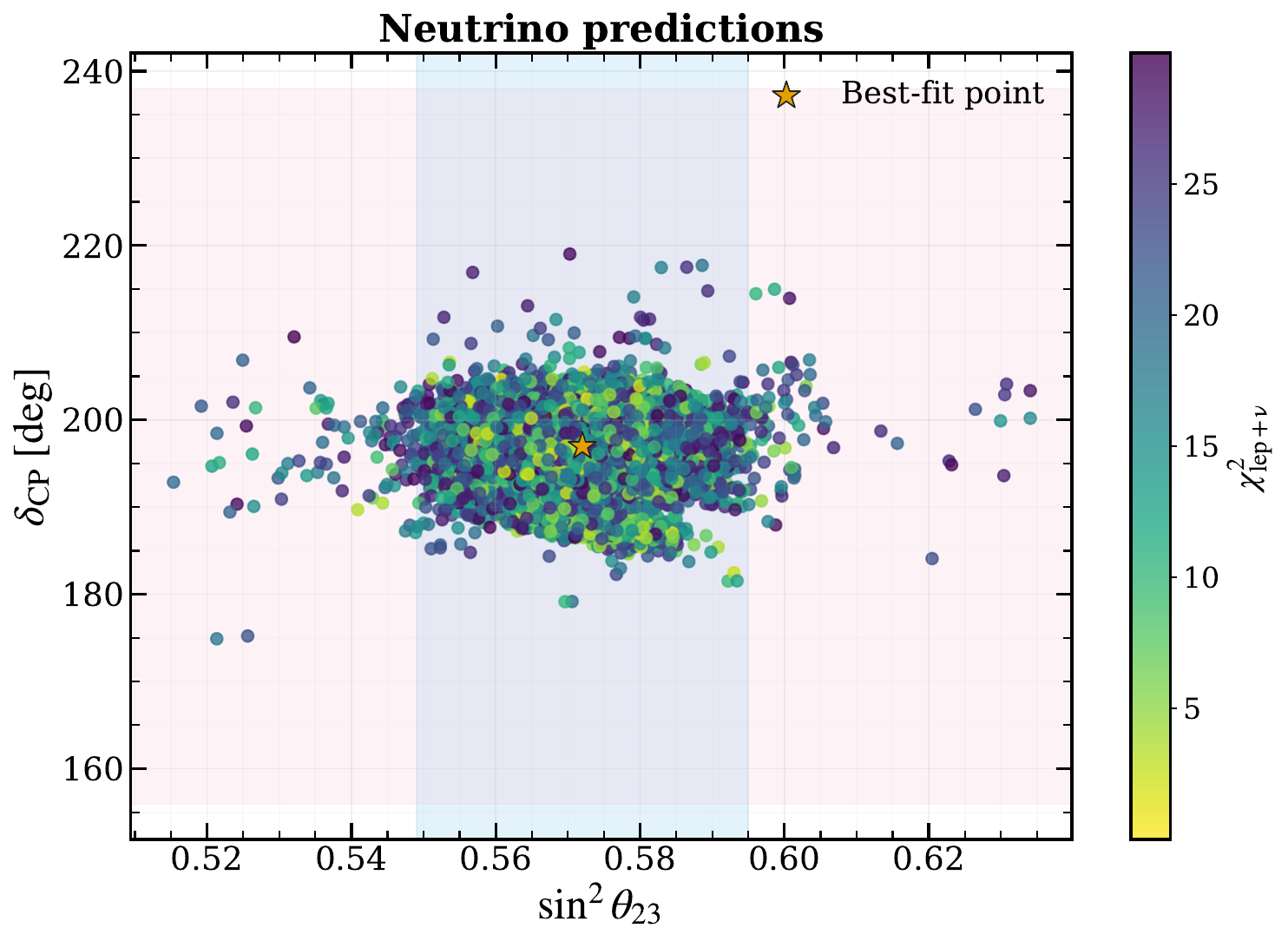}
\hfill
\includegraphics[width=0.48\linewidth]{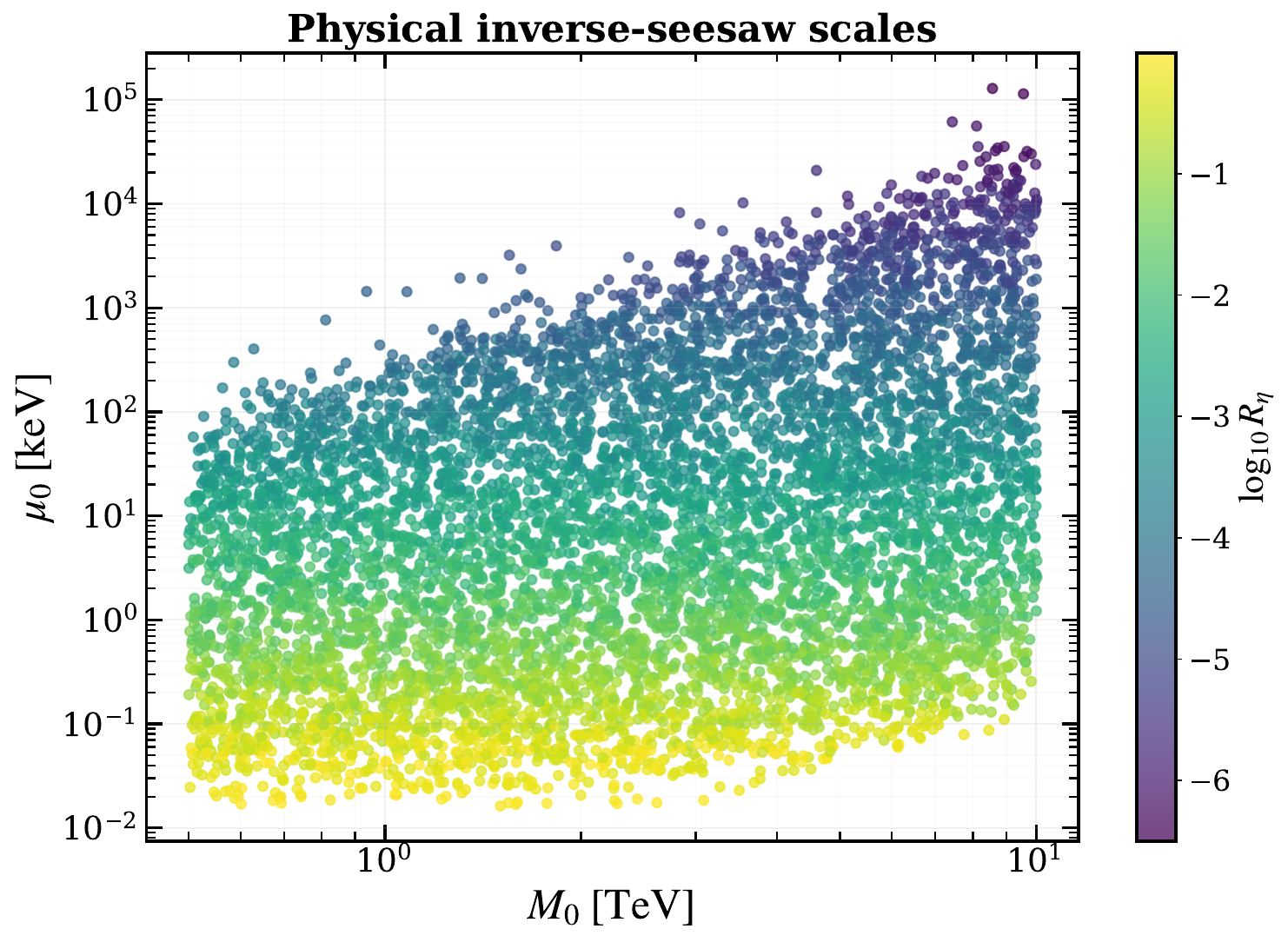}
\caption{Lepton and inverse-seesaw predictions.  Left: accepted points in
the $(\sin^2\theta_{23},\delta_{\rm CP})$ plane, colored by the total
lepton-sector fit quality.  The concentration of points around the
preferred region is a nontrivial consequence of the common modular
texture and the charged-lepton rotation.  Right: physical inverse-seesaw
scales after restoring $(c_0,M_0,\mu_0)$; the color gives the normalized
non-unitarity measure $R_\eta$.  The scan admits TeV-scale heavy states
while maintaining small light-neutrino masses through the inverse-seesaw
suppression.}
\label{fig:nu-scale}
\end{figure}

Figure~\ref{fig:nu-scale} (left) shows the correlation between
$\sin^2\theta_{23}$ and $\delta_{\rm CP}$.  The accepted cloud is compact
despite the high-dimensional scan, demonstrating that the common modular
texture constrains the atmospheric angle and CP phase rather than merely
fitting them independently.  The color encodes the full
$\chi^2_{\ell+\nu}$ and shows that the best-fit region lies inside the
current oscillation bands.

\begin{figure}[t]
\centering
\includegraphics[width=0.78\linewidth]{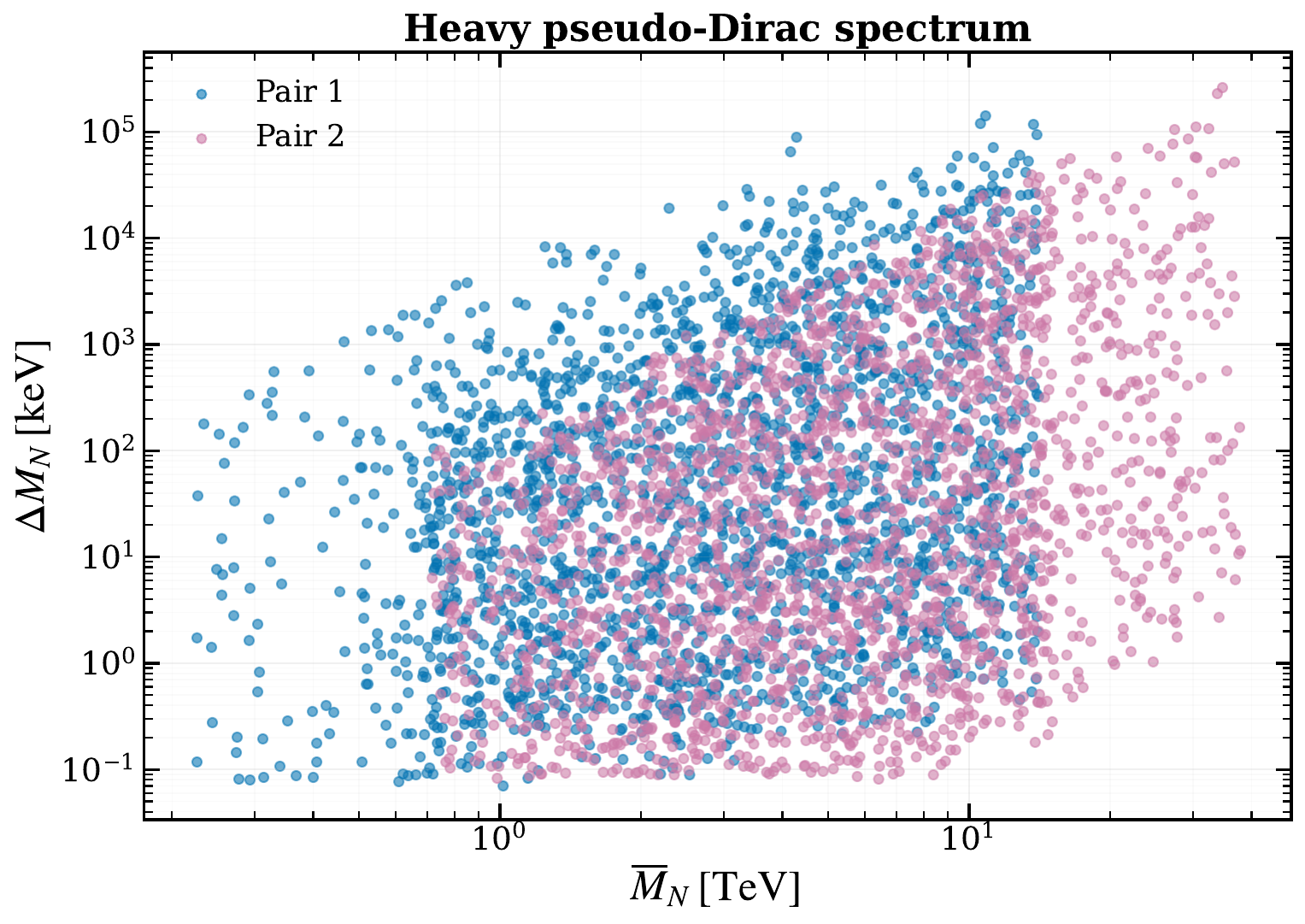}
\caption{Heavy inverse-seesaw spectrum.  Each point represents one of the
two quasi-Dirac pairs, plotted as its average mass versus the Majorana
splitting.  The approximate lepton-number symmetry of the inverse seesaw
allows the heavy masses to remain near the TeV scale while the splittings
are controlled by the small sterile Majorana block $\mu_S$.}
\label{fig:heavy}
\end{figure}

The rank-two prediction gives a negligible lightest mass.  A representative
benchmark has
\begin{equation}
(m_1,m_2,m_3)=
(1.76\times10^{-10},\,8.608\times10^{-3},\,5.01099\times10^{-2})\,\eV,
\end{equation}
and
\begin{equation}
\sum_i m_i=0.05872\,\eV,
\label{eq:sum}
\end{equation}
comfortably below the characteristic cosmological scale probed by current
datasets \cite{Planck2018}.
The absolute inverse-seesaw scale is restored by scanning $c_0$ and $M_0$
and solving Eq.~\eqref{eq:kappanu} for
\begin{equation}
\mu_0=
\frac{\kappa_\nu M_0^2}{v_u^2c_0^2}.
\label{eq:mu0}
\end{equation}
The right panel of Fig.~\ref{fig:nu-scale} displays the resulting
$(M_0,\mu_0)$ space.  The broad diagonal structure is not accidental:
holding the light mass scale fixed requires
$\mu_0\propto M_0^2/c_0^2$.  The color is the normalized non-unitarity
measure $R_\eta$ from Eq.~\eqref{eq:Reta}.  Of the $40000$ scale points,
$34807$ pass the element-by-element filter; their median normalized
non-unitarity is $R_\eta=4.31\times10^{-3}$.
In our scan, the physical-scale layer contains
$2\times10^5$ points; only points satisfying $R_\eta<1$ are displayed in
the physical inverse-seesaw parameter space.
The physical inverse-seesaw sample organizes the four heavy Majorana states
into two quasi-Dirac pairs.  Figure~\ref{fig:heavy} displays this behavior
over the viable scale sample: the average pair masses extend from the
sub-TeV to multi-TeV regime, while the pair splittings are controlled by
the small sterile Majorana matrix $\mu_S$ and its modular texture.

\section{Common-modulus dark parameter space}

The dark scan is performed only after the lepton-sector acceptance step.
For each dark point we first draw
\begin{equation}
\tau\in\{\tau_{\rm ISS}\},
\end{equation}
from the accepted inverse-seesaw sample, evaluate the same modular forms,
and then scan
\begin{equation}
10^{-3}\le |y_D|,|y_M|\le3,
\quad
10^8\GeV\le v_\Phi\le10^{11}\GeV,
\quad
10^{14}\GeV\le\Lambda\le10^{18}\GeV,
\label{eq:darkranges}
\end{equation}
with $\Lambda>v_\Phi$. In our scan, the common-modulus dark sector contains
$N_{\rm dark}=2\times10^6$ sampled points. We  define
the first LZ-oriented region by
\begin{equation}
300\GeV\le m_\chi\le2000\GeV,
\qquad
300\keV\le\delta\le400\keV.
\label{eq:LZbox}
\end{equation}
This yields
\begin{equation}
N_{\rm ISS+LZ}=905,
\qquad
\epsilon_{\rm scan}=0.3017\%.
\label{eq:darkacc}
\end{equation}
The percentage is a Monte-Carlo scan acceptance fraction, not a
statistical confidence level.

\begin{figure}[t]
    \centering

    \begin{subfigure}{0.6\linewidth}
        \centering
        \includegraphics[width=\linewidth]{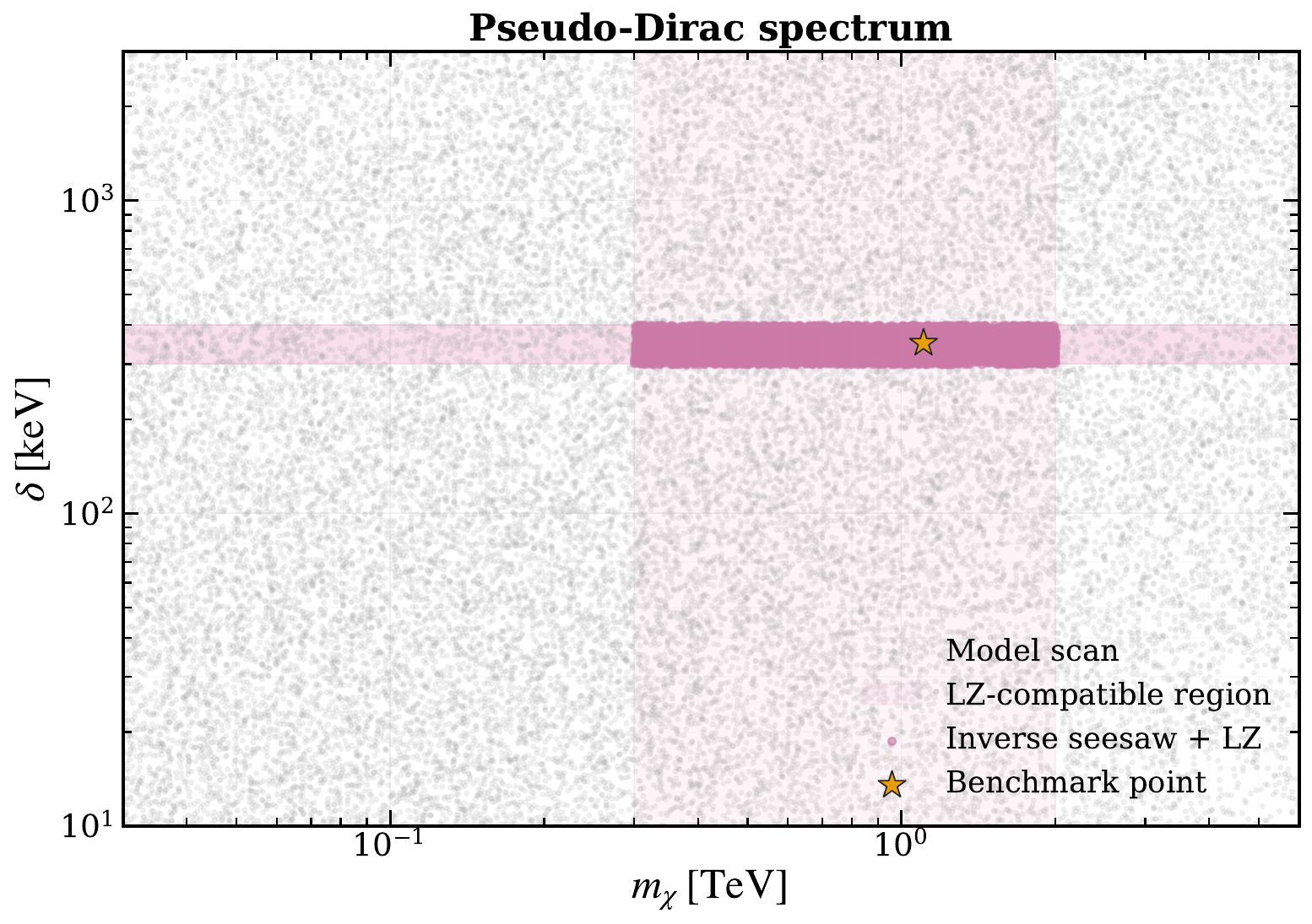}
        \caption{Dark-sector spectrum.}
        \label{fig:dark-spectrum}
    \end{subfigure}

    \vspace{0.5cm}

    \begin{subfigure}{0.9\linewidth}
        \centering
        \includegraphics[width=\linewidth]{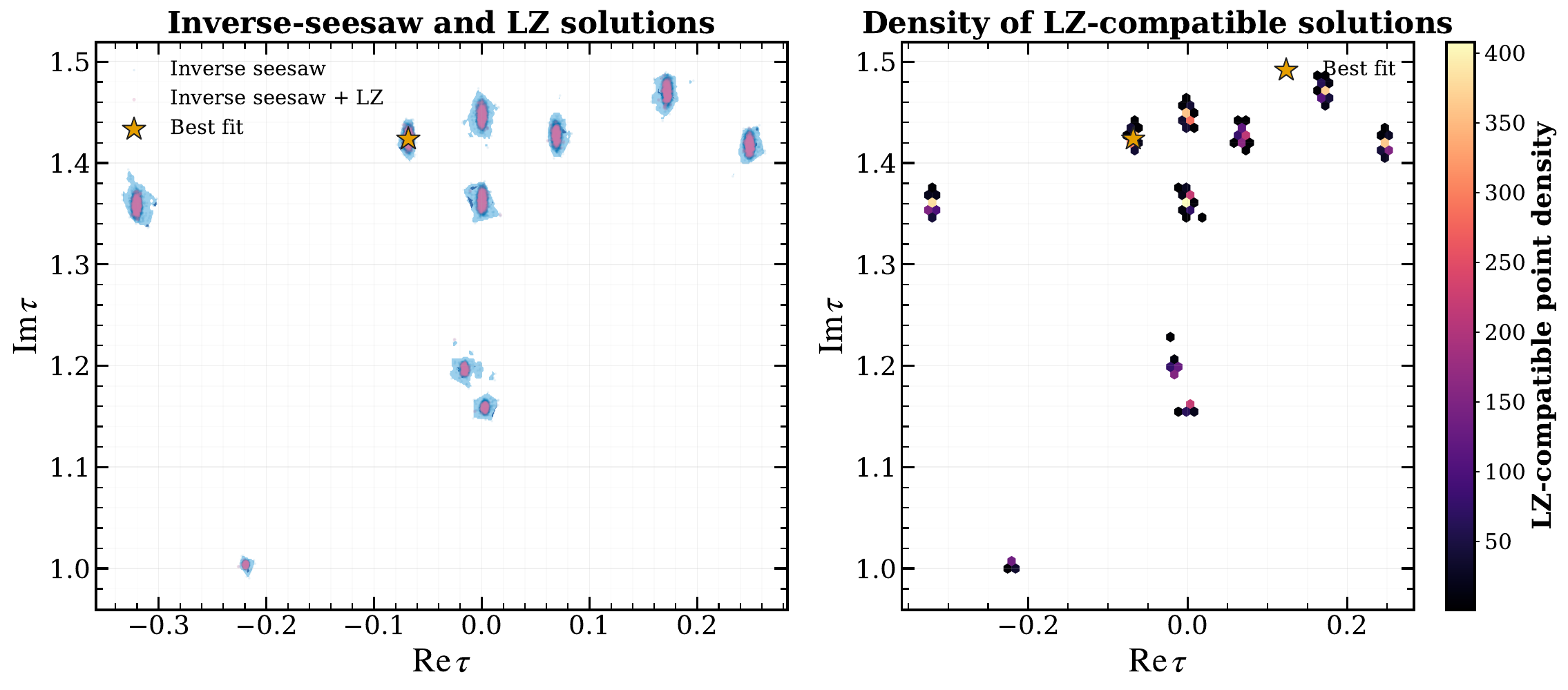}
        \caption{Common modulus plane for the accepted dark points.}
        \label{fig:common-modulus}
    \end{subfigure}

    \caption{From modular flavor to the LZ-oriented dark spectrum.
    \textbf{Top:} model-generated pseudo-Dirac masses and splittings.
    Magenta points satisfy
    $300\GeV\le m_\chi\le2\TeV$ and
    $300\keV\le\delta\le400\keV$.
    \textbf{Bottom:} the same accepted dark points shown in the modular
    plane. Their localization within the inverse-seesaw islands demonstrates
    that the dark scan is not independent of the flavor fit: the same modulus
    controls both sectors.}
    \label{fig:dark-common}
\end{figure}

Figure~\ref{fig:dark-common} (left) shows how the modular operators populate
the $(m_\chi,\delta)$ plane.  The gray cloud is the wider model scan,
whereas magenta points also satisfy Eq.~\eqref{eq:LZbox}.  The accepted
points form a narrow horizontal band in the few-hundred-keV splitting
region while covering a broad TeV-scale mass range.  This is the direct
numerical realization of Eq.~\eqref{eq:ratio}: once $\tau$ fixes the ratio
of modular forms, the hierarchy $\delta/m_\chi$ is controlled by
$v_\Phi/\Lambda$ and the ratio $y_M/y_D$.

The right panel of Fig.~\ref{fig:dark-common} makes the common-modulus
condition explicit.  The magenta inverse-seesaw+LZ points do not populate
the entire modular plane; they inherit the disconnected flavor islands.
The density panel shows that some portions of those islands are much more
efficient at producing the desired pseudo-Dirac spectrum.  This is the
central model correlation of the paper: the LZ-relevant dark parameters
are filtered by the same $\tau$ that already passed charged-lepton and
neutrino data.

\begin{figure}[t]
\centering
\includegraphics[width=0.76\linewidth]{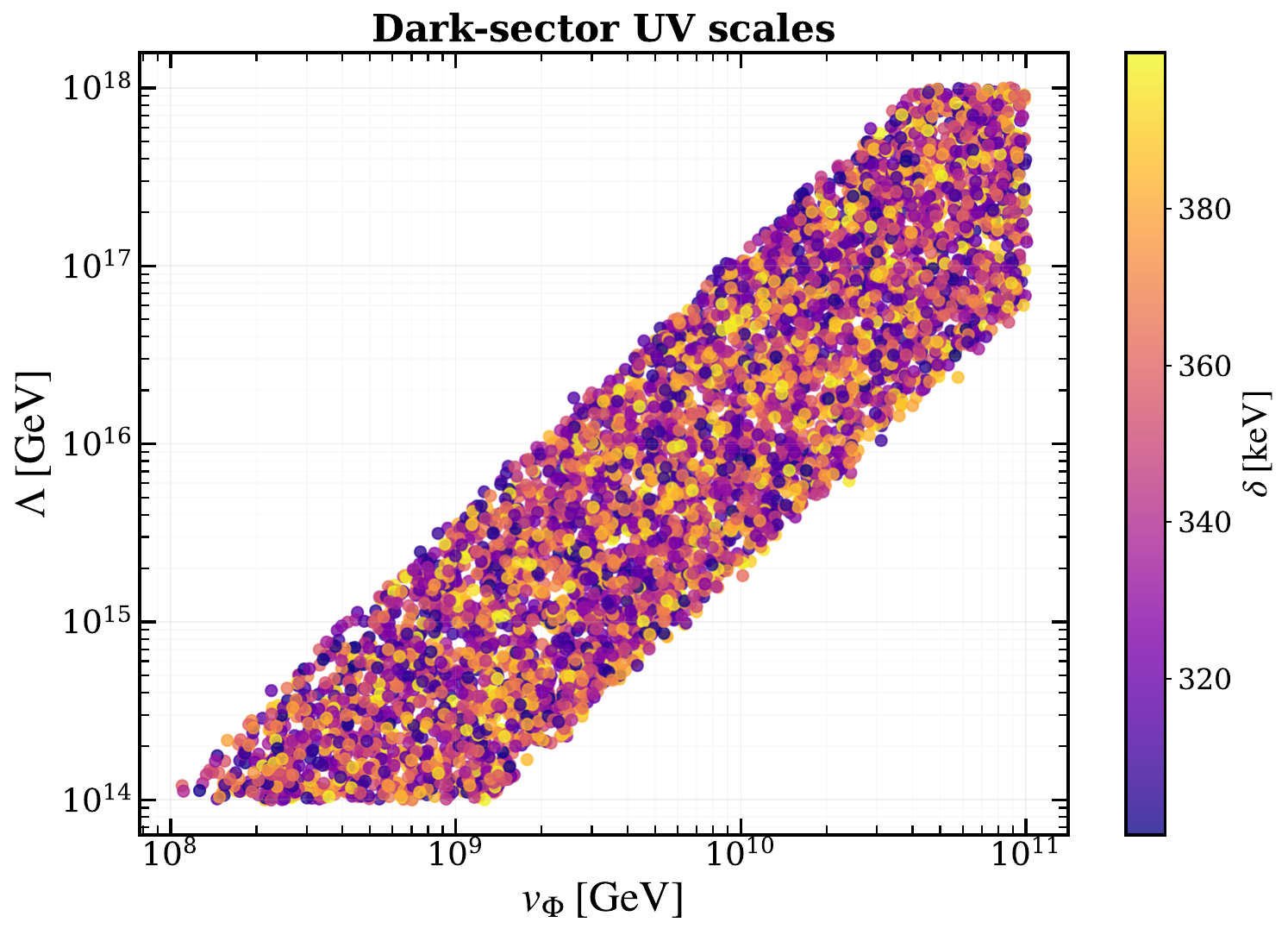}
\caption{Ultraviolet scales for the common-modulus dark solutions.  The
accepted points form a correlated band in $(v_\Phi,\Lambda)$, with color
indicating the inelastic splitting.  The band follows from the simultaneous
requirements on the TeV Dirac mass and the much smaller Majorana
perturbation, while its width reflects the scanned coefficients and the
modular-form ratios.}
\label{fig:uv}
\end{figure}

The ultraviolet interpretation is shown in Fig.~\ref{fig:uv}.  Accepted
points follow a strong positive correlation between $v_\Phi$ and $\Lambda$.
For fixed $m_\chi$ and $\delta$, Eqs.~\eqref{eq:UVexact} predict
$\Lambda\propto v_\Phi^2/m_\chi$ together with the modular dependence of
$A$ and $B$.  The color variation with $\delta$ shows how the
few-hundred-keV splitting selects a relatively narrow hierarchy of
high scales.  For order-one modular coefficients, a TeV dark mass and a
few-hundred-keV splitting correspond parametrically to
$v_\Phi\sim10^9\GeV$ and a cutoff near the conventional unification regime,
although the exact location is shifted by $A(\tau)$ and $B(\tau)$.

\section{LZ high-recoil kinematics and likelihood}

For endothermic scattering on a nucleus of mass $m_N$,
\begin{equation}
v_{\min}(E_R)=
\frac{1}{\sqrt{2m_NE_R}}
\left(
\frac{m_NE_R}{\mu_{\chi N}}+\delta
\right),
\qquad
\mu_{\chi N}=\frac{m_\chi m_N}{m_\chi+m_N}.
\label{eq:vmin}
\end{equation}
The derivation is given in Appendix~\ref{app:kin}.  A terrestrial
transition is possible only if
\begin{equation}
\delta\le
\delta_{\max}
=
\frac12\mu_{\chi N}v_{\max}^2,
\label{eq:dmax}
\end{equation}
and the kinematically allowed recoil interval is
\begin{equation}
E_R^\pm=
\frac{\mu_{\chi N}^2}{2m_N}
\left[
v_{\max}\pm
\sqrt{v_{\max}^2-\frac{2\delta}{\mu_{\chi N}}}
\right]^2.
\label{eq:ERpm}
\end{equation}
We use the Standard Halo Model reference values
\begin{equation}
v_0=220~{\rm km\,s^{-1}},
\qquad
v_E=220~{\rm km\,s^{-1}},
\qquad
v_{\rm esc}=544~{\rm km\,s^{-1}},
\end{equation}
so $v_{\max}=764~{\rm km\,s^{-1}}$.  Direct-detection conventions,
nuclear responses and the Helm form factor follow the standard literature
\cite{Lewin:1995rx,Lewin:1995rx,FitzpaLewin:1995rxtrickEtAl2013,
AnandFitzpatrickHaxton2014}.

\begin{figure}[t]
\centering
\includegraphics[width=0.48\linewidth]{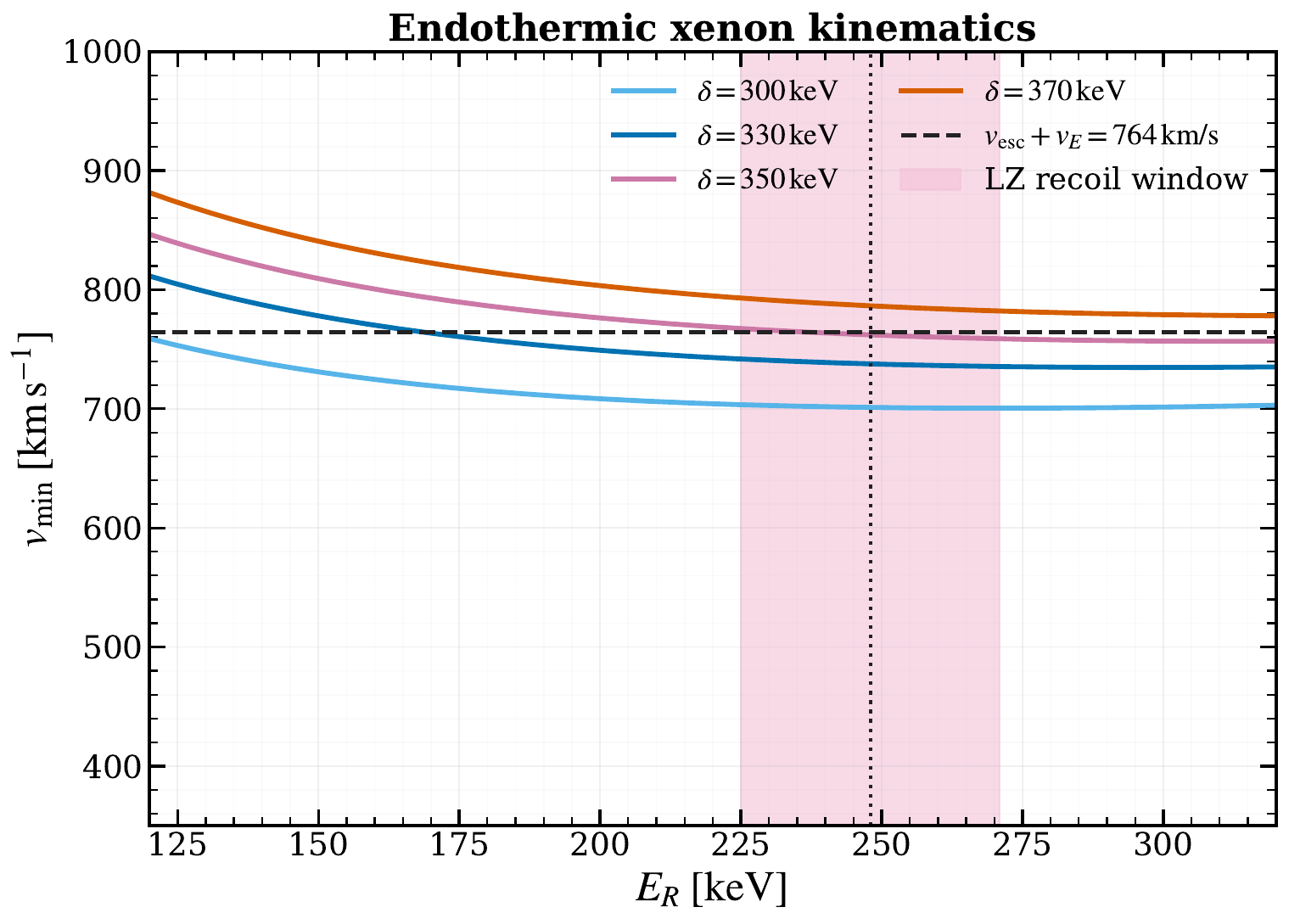}
\hfill
\includegraphics[width=0.48\linewidth]{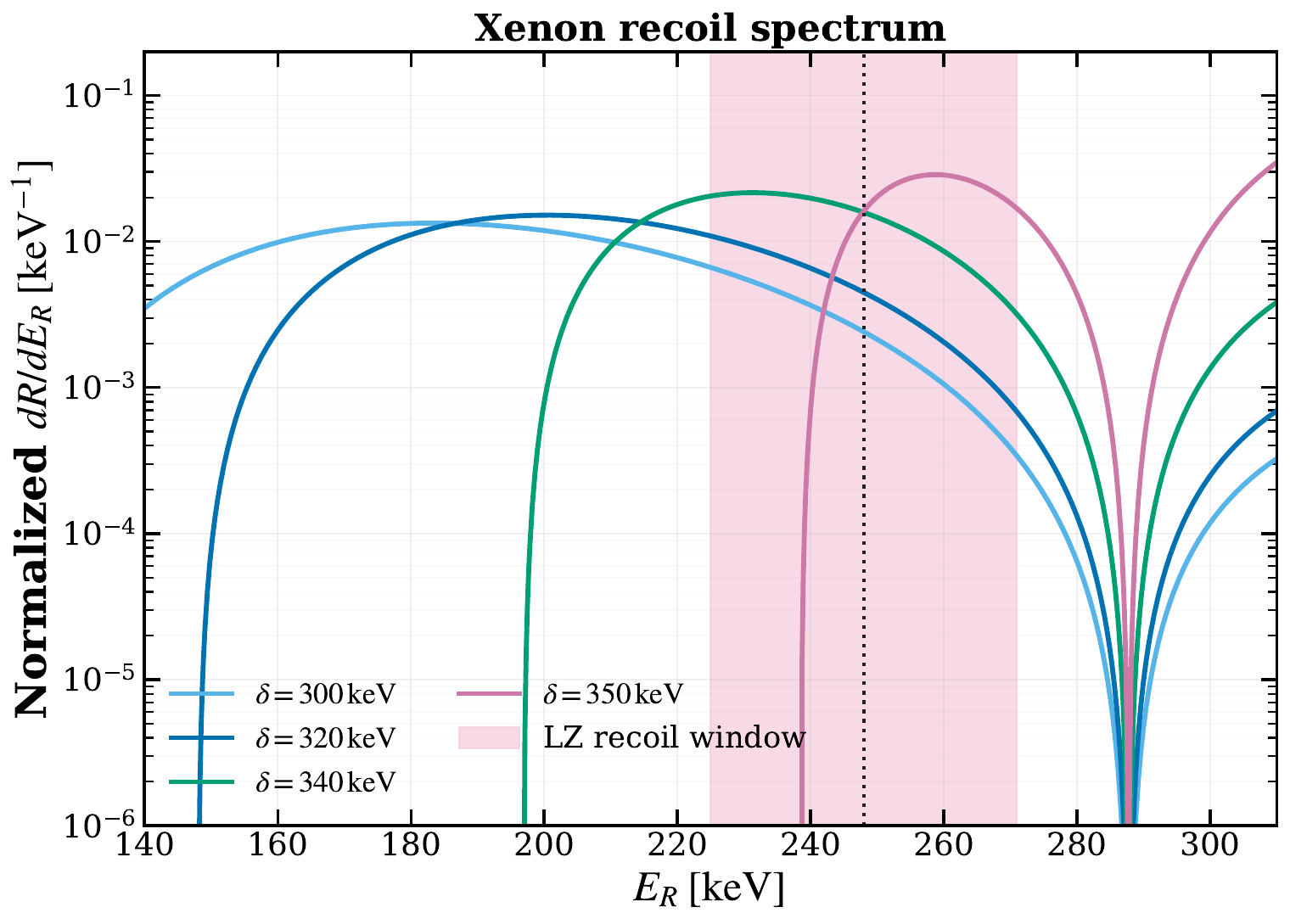}
\caption{Endothermic xenon kinematics.  Left: minimum incoming speed for
representative splittings, with the LZ $225$--$271\keV$ interval shaded
and the $248\keV$ central event energy marked.  The few-hundred-keV
splitting pushes the scattering into the extreme Galactic velocity tail.
Right: normalized xenon recoil shapes.  Increasing $\delta$ removes the
low-energy spectrum and shifts support toward the high-recoil window;
the high-energy dip and recovery are generated by the xenon Helm form
factor.}
\label{fig:kinematics}
\end{figure}

Figure~\ref{fig:kinematics} (left) shows $v_{\min}$ for representative
splittings.  The LZ interval $225$--$271\keV$ lies precisely where the
$\delta\sim350\keV$ curve approaches the maximum Galactic speed.  The
kinematic interpretation is therefore sharp: the high-recoil event probes
the last part of the terrestrial phase space before endothermic scattering
closes.  Increasing $\delta$ further eliminates the process unless either
$m_\chi$ is increased or one invokes a larger high-speed halo tail.

At shape level,
\begin{equation}
\frac{dR}{dE_R}\propto
F_{\rm Helm}^2(E_R)\,
\eta[v_{\min}(E_R)],
\label{eq:shape}
\end{equation}
where $\eta$ is the mean inverse speed.  The right panel of
Fig.~\ref{fig:kinematics} shows normalized xenon spectra.  The low-energy
support disappears as $\delta$ rises, while the surviving rate migrates
toward the LZ high-recoil interval.  The pronounced structure above the
event region comes from the xenon form factor, not from a resonance in the
dark sector.  This point is especially relevant because recent
model-independent work has emphasized the role of sidebands and the xenon
diffraction structure when assessing inelastic interpretations
\cite{PalmisanoTammaroTesi2026,DelepineKhalil2026}.

The experimental energy resolution is essential near such a sharp
kinematic endpoint.  Combining the quoted statistical and systematic
uncertainties in quadrature gives
\begin{equation}
\sigma_E=
\sqrt{(23\keV)^2+(23\keV)^2}
=32.53\keV.
\label{eq:sigmaE}
\end{equation}
The true spectrum is therefore convolved with
\begin{equation}
\left(\frac{dR}{dE_R}\right)_{\rm sm}
=
\int dE_R'\,
\frac{dR}{dE_R'}
\frac{
\exp[-(E_R-E_R')^2/(2\sigma_E^2)]
}{
\sqrt{2\pi}\sigma_E
}.
\label{eq:smear}
\end{equation}
Smearing moves spectral weight across the nominal endpoint and is
particularly important for solutions lying close to Eq.~\eqref{eq:dmax},
as also found in recent high-recoil model studies
\cite{Yamashita2026,DiMauro2026,SuYangYang2026}.

To go beyond the rectangular selection in Eq.~\eqref{eq:LZbox}, we construct
a deliberately transparent two-bin Poisson likelihood using
\begin{equation}
B_1=[160,200]\keV,
\qquad
B_2=[225,271]\keV,
\qquad
(n_1,n_2)=(0,1),
\label{eq:bins}
\end{equation}
following the illustrative strategy of Ref.~\cite{SuYangYang2026}.  The
smeared shape is integrated over each bin,
\begin{equation}
S_i(m_\chi,\delta)=
\int_{B_i}dE_R
\left(\frac{dR}{dE_R}\right)_{\rm sm},
\end{equation}
and the expected signal is
\begin{equation}
\lambda_i=\mu S_i,
\end{equation}
where $\mu\ge0$ is an unknown total normalization.  We profile over $\mu$.
For independent Poisson bins,
\begin{equation}
\ln\mathcal L=
\sum_i
\left[
n_i\ln\lambda_i-\lambda_i-\ln(n_i!)
\right],
\end{equation}
and define
\begin{equation}
\Delta\chi^2(m_\chi,\delta)=
-2\left[
\ln\mathcal L_{\rm prof}(m_\chi,\delta)
-\ln\mathcal L_{\rm max}
\right].
\label{eq:dchi2}
\end{equation}
The analytic profiling relation is derived in
Appendix~\ref{app:likelihood}.  We display the conventional two-parameter
reference values
\begin{equation}
\Delta\chi^2=2.30,\qquad5.99,\qquad11.83,
\end{equation}
as $68\%$, $95\%$, and $3\sigma$ guides.  Because this problem contains
one event, a bounded normalization, kinematic edges, and no full LZ
background-nuisance model, these should not be interpreted as exact
frequentist coverage.

\begin{figure}[t]
\centering
\includegraphics[width=0.88\linewidth]{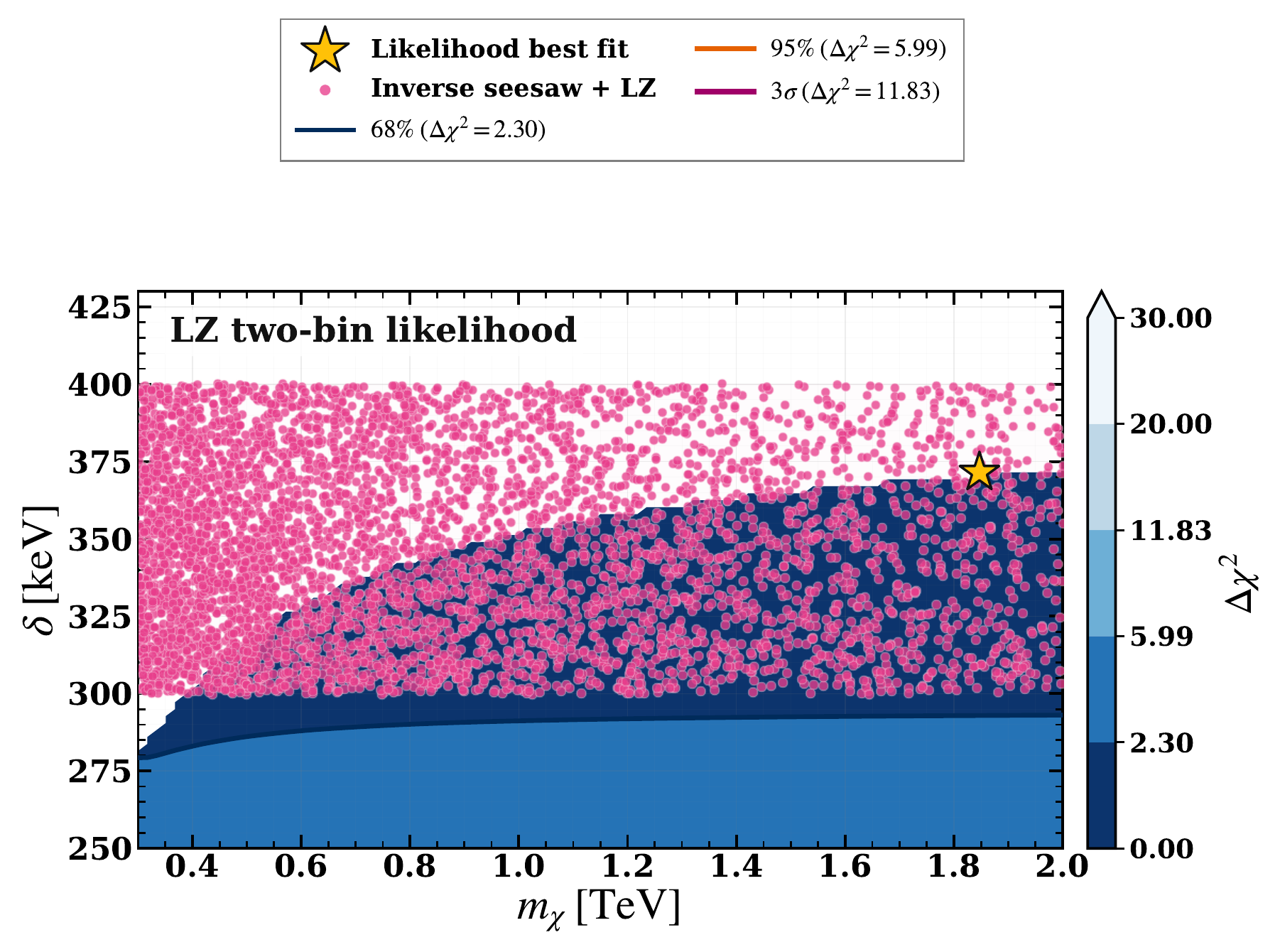}
\caption{Exploratory LZ two-bin likelihood in the
$(m_\chi,\delta)$ plane.  Darker blue denotes smaller $\Delta\chi^2$.
The magenta points are common-modulus inverse-seesaw+LZ solutions and the
gold star is the likelihood best grid point,
$(1.864\TeV,371.5\keV)$.  The reference contours use the conventional
two-parameter values $\Delta\chi^2=2.30$, $5.99$, and $11.83$.  The upper
preferred boundary closely follows the endothermic xenon kinematic ceiling.
This figure is a spectral diagnostic and does not reproduce the official
LZ detector likelihood.}
\label{fig:likelihood}
\end{figure}

The likelihood is evaluated on a $101\times81$ grid spanning
$300\le m_\chi/{\rm GeV}\le2000$ and
$250\le\delta/{\rm keV}\le430$.  The best grid point is
\begin{equation}
m_\chi^{\rm bf}=1.847\TeV,
\qquad
\delta^{\rm bf}=371.5\keV.
\label{eq:LZbest}
\end{equation}
For xenon with $A=131$ and $v_{\max}=764~{\rm km\,s^{-1}}$,
Eq.~\eqref{eq:dmax} gives
\begin{equation}
\delta_{\max}(1.847\TeV)\simeq371.69\keV.
\label{eq:dmaxbf}
\end{equation}
The best fit therefore sits almost exactly on the terrestrial inelastic
ceiling.  This explains the likelihood geometry in
Fig.~\ref{fig:likelihood}: for each mass, the preferred upper boundary is
set by the largest splitting that still allows the high-recoil event,
while detector smearing permits endpoint spectral weight to migrate into
the measured interval.  The magenta model points demonstrate that the
common-modulus construction populates this kinematically selected region.

\textcolor{black}{The likelihood should also be interpreted in light of the rapidly
evolving comparison between LZ and solar-neutrino searches.  Several
analyses find that endothermic solutions are selected close to a
terrestrial kinematic boundary and are consequently sensitive to the
high-speed tail of the Galactic halo
\cite{DiMauro2026,SuYangYang2026,PalmisanoTammaroTesi2026,
DelepineKhalil2026}.  For the conventional nearly pure thermal Higgsino
near $m_\chi\simeq1.1\TeV$, solar capture followed by annihilation into
electroweak final states can generate a high-energy neutrino signal that is
strongly constrained by IceCube and Super-Kamiokande
\cite{PospelovRamani2026,BoseEtAlHiggsino2026}.  This places substantial
pressure on the simplest thermal-Higgsino interpretation of the LZ event,
but it does not exclude the Higgsino framework as a whole.}

\textcolor{black}{The situation changes significantly for heavier Higgsino dark matter.
Dedicated analyses demonstrate that the solar-neutrino constraint can be
substantially relaxed while maintaining an LZ-compatible inelastic
splitting.  Ref.~\cite{Langhoff2026} finds viable very-heavy-Higgsino
solutions after combining the LZ high-energy sideband with solar-capture
information, while Ref.~\cite{FreeseTheodosopoulos2026} identifies joint
LZ--IceCube compatible regions extending from masses of order a few TeV
to much larger values, depending on the assumed high-velocity halo
distribution, solar cooling processes, and sideband treatment.
Heavy-Higgsino scenarios with nonstandard cosmological histories provide
an additional realization of this possibility
\cite{Pallis2026HeavyHiggsino}.  The IceCube result should therefore be
interpreted as a powerful discriminator of the Higgsino mass and thermal
history, rather than as a universal exclusion of inelastic Higgsino dark
matter.}

\textcolor{black}{This distinction is particularly relevant for the present construction.
Our dark state is a Standard-Model singlet rather than an electroweak
Higgsino, and its solar capture rate is controlled by the model-dependent
vector portal of Eq.~\eqref{eq:portal}.  Higgsino-specific solar-neutrino
limits therefore cannot be transferred directly to the modular singlet
model.  At the same time, the heavy-Higgsino studies provide an important
proof of principle that LZ high-recoil kinematics and IceCube constraints
can coexist in suitable regions of mass, splitting, halo assumptions, and
cosmological parameter space.  A dedicated solar-capture calculation for
the present portal would thus constitute a complementary probe rather than
an automatic exclusion.}

The model contains two distinct pseudo-Dirac structures, and keeping them
separate is essential.  The heavy inverse-seesaw neutral fermions are
quasi-Dirac because the sterile Majorana matrix $\mu_S$ softly violates
lepton number.  The dark pair $\chi_{1,2}$ is independently pseudo-Dirac,
with a TeV-scale Dirac mass and a few-hundred-keV Majorana perturbation.
The numerical values of $\mu_0$ and $\delta$ can both lie in the broad keV
domain, but they belong to different matrices and should not be identified.
Their correlation arises only through the common modular background.

This architecture distinguishes the present construction from recent
neutrino-linked LZ models.  The $A_4$ modular Dirac inverse seesaw of
Ref.~\cite{DasKarmakarMahapatraPaul2026}, the type-II-assisted
singlet-doublet model of Ref.~\cite{BorahSingletDoublet2026}, the radiative
Dirac model of Ref.~\cite{BorahRadiativeDirac2026}, and the radiative
seesaw construction of Ref.~\cite{OkadaShigekamiWu2026} all correlate
neutrino physics and inelastic scattering, but through different symmetry
and mediator structures.  Here the characteristic correlation is instead
the common value of $\tau$ entering both the inverse-seesaw flavor texture
and the higher-dimensional dark mass operators.

The ultraviolet scale relation in Eq.~\eqref{eq:UVexact} is also
phenomenologically useful.  A few-hundred-keV splitting need not be inserted
as a tiny bare parameter.  It follows from a ratio of symmetry-breaking
and cutoff scales, multiplied by modular functions and order-one or
moderately hierarchical couplings.  This logic is complementary to
geometric suppression in extra-dimensional constructions
\cite{AhmedLeontaris2026,LeeRandall2026}, loop-induced splittings
\cite{OkadaShigekamiWu2026}, and mixing suppression
\cite{LeeYoun2026}.

A particularly attractive feature of the present construction is that the
LZ-motivated hierarchy does not have to be introduced as an isolated
low-energy tuning.  The TeV-scale pseudo-Dirac mass and the
few-hundred-keV Majorana splitting originate from operators of different
dimension,
\begin{equation}
M_\chi\propto
Y_{\mathbf 1}^{(4)}(\tau)\frac{v_\Phi^2}{\Lambda},
\qquad
\mu_\chi\propto
Y_{\mathbf 1}^{(6)}(\tau)\frac{v_\Phi^3}{\Lambda^2},
\end{equation}
so that the small ratio $\delta/m_\chi$ is controlled by the hierarchy
$v_\Phi/\Lambda$, together with calculable modular functions and
dimensionless couplings.  Consequently, the sub-MeV splitting required by
the LZ high-recoil kinematics can be traced to the ultraviolet structure of
the theory rather than inserted by hand.  For order-one combinations of
couplings and modular forms, the preferred TeV/few-hundred-keV region
naturally points toward an intermediate symmetry-breaking scale and a
cutoff extending toward the conventional grand-unification regime.  The
numerical $(v_\Phi,\Lambda)$ correlation shown above is therefore not merely
a consequence of parameter scanning, but a direct reflection of the
operator hierarchy in the underlying theory.

The common-modulus structure provides an additional source of predictivity.
The dark sector is not scanned independently over the full complex
$\tau$ plane; instead, it inherits only those modular configurations that
have already reproduced the charged-lepton hierarchy and neutrino
oscillation data.  In this sense, the LZ-motivated pseudo-Dirac spectrum is
filtered by flavor physics.  The fact that viable
$(m_\chi,\delta)$ points survive this restriction demonstrates that the
high-recoil region can be embedded consistently into the same modular
background that governs the inverse seesaw.  At the same time, the
inverse-seesaw and dark pseudo-Dirac sectors remain physically distinct,
so that the small lepton-number-violating scale responsible for neutrino
masses need not be artificially identified with the dark-matter splitting.
This separation preserves the successful inverse-seesaw mechanism while
allowing the dark sector to realize the much heavier state required for
xenon scattering.

The framework is also naturally suited to several independent
phenomenological tests.  A specified vector portal would convert the
spectral analysis presented here into an absolute-rate prediction and
would connect the LZ signal region with relic abundance, accelerator
searches, and solar capture
\cite{FabbrichesiGabrielliLanfranchi2020,LeeBaryonFlavor2026,
NguyenLindenHooper2026}.  Depending on the mediator and cosmological
history, the abundance may arise through annihilation, coannihilation,
freeze-in, or nonthermal production, mechanisms already appearing in
recent LZ-motivated constructions
\cite{DiMauro2026,LeeBaryonFlavor2026,
HeHongJeesun2026,ChoudhuryDasSachdeva2026}.  These possibilities enlarge,
rather than weaken, the phenomenological reach of the modular framework,
because the flavor prediction and the origin of the pseudo-Dirac splitting
are largely independent of the particular mechanism that sets the relic
density.

The heavy-Higgsino literature provides a useful parallel in this
respect.  Once the mass is raised substantially above the conventional
thermal-Higgsino scale, modified cosmological histories such as early
matter domination or low-temperature reheating can recover the observed
dark-matter abundance while retaining the inelastic spectrum relevant for
the LZ event \cite{Langhoff2026,Pallis2026HeavyHiggsino}.  This reinforces
the separation adopted here between the modular origin of the
pseudo-Dirac spectrum and the later choice of the cosmological production
mechanism.

Future high-recoil data can test the construction in several complementary
ways.  Increased LZ exposure will determine whether the present event is
reinforced by additional recoils, while the spectral sidebands can
discriminate an endothermic kinematic edge from exothermic or
momentum-dependent elastic interpretations
\cite{PalmisanoTammaroTesi2026,DelepineKhalil2026,BaerBarger2026}.
Moreover, different nuclei probe different reduced masses, inelastic
thresholds, and form-factor structures.  Multi-target measurements with
xenon, tungsten, argon, and other nuclei therefore provide a particularly
powerful test of the pseudo-Dirac interpretation
\cite{SuYangYang2026,AhmedLeontaris2026,AhmedAhmadRehman2026}.
The isotope dependence near the xenon kinematic edge offers an additional
handle that is absent in conventional low-recoil searches.

Taken together, these features make the construction more than a
phenomenological fit to a single recoil event.  The same modulus controls
the lepton flavor structure and participates in generating the dark
pseudo-Dirac hierarchy; the inverse seesaw naturally accommodates
TeV-scale heavy neutral fermions while keeping the active-neutrino masses
small; and the dark-sector splitting has a transparent ultraviolet origin
through the ratio of symmetry-breaking and cutoff scales.  The
LZ high-recoil region therefore emerges as a low-energy probe of a much
broader structure linking modular flavor, neutrino mass generation,
pseudo-Dirac dark matter, and ultraviolet physics.

\section{Conclusions}

We have constructed a modular-$S_4$ framework that connects lepton flavor
to the LZ high-recoil regime without identifying the neutrino singlets with
the heavy dark particle.  The charged-lepton sector retains all independent
$S_4$ contractions, while a minimal $2N_R+2S$ inverse seesaw generates a
rank-two light-neutrino matrix and TeV-scale heavy quasi-Dirac states.Our scan yields $181447$ accepted solutions
with $\chi^2_{\ell+\nu}<30$, resolving ten distinct locally refined
modular basins with both positive- and negative-${\rm Re}\,\tau$ branches.
A separate singlet pair acquires a pseudo-Dirac spectrum from
higher-dimensional modular operators.  The same $\tau$ selected by the
lepton fit enters the dark Dirac mass and Majorana splitting.
In our scan, the common-modulus dark sector contains $2\times10^6$
sampled points, with the LZ-oriented TeV/few-hundred-keV subset used
for the spectral and modular-plane overlays.  Their ultraviolet scales
form a correlated band that follows analytically from the hierarchy between the dimension-five and dimension-six mass operators.
The high-recoil phenomenology is controlled by endothermic kinematics.
The few-hundred-keV splitting pushes xenon scattering toward the maximum
terrestrial halo speed and removes much of the conventional low-energy
spectrum.  Using our $101\times81$ likelihood grid, the best grid point is
$(m_\chi,\delta)=(1.847\TeV,371.5\keV)$, approximately $0.19\keV$ below
the xenon kinematic ceiling for the adopted halo parameters.  This provides a
simple physical interpretation of the likelihood boundary and shows that
the modular model naturally populates the relevant region.

\appendix

\section{Charged-lepton contractions and normalization}
\label{app:charged}

For two $S_4$ doublets
$A=(A_1,A_2)^T$ and $B=(B_1,B_2)^T$, the contractions used in the main
text are
\begin{align}
(AB)_{\mathbf1}&=A_1B_1+A_2B_2,\nonumber\\
(AB)_{\mathbf1'}&=A_1B_2-A_2B_1,\nonumber\\
(AB)_{\mathbf2}&=
\begin{pmatrix}
A_2B_2-A_1B_1\\
A_2B_1+A_1B_2
\end{pmatrix}.
\end{align}
Using $A=e_D^c$, $B=L_D$, and
$Y_{\mathbf2}^{(6)}=S_Y(Y_1,Y_2)^T$ gives
\begin{align}
[(e_D^cL_D)_{\mathbf2}Y_{\mathbf2}^{(6)}]_{\mathbf1}
={}&
S_Y(e_2^cL_2-e_1^cL_1)Y_1
\nonumber\\
&+
S_Y(e_2^cL_1+e_1^cL_2)Y_2,
\end{align}
which directly yields Eq.~\eqref{eq:Yell} together with the singlet and
primed-singlet contractions.

Writing
\begin{equation}
Y_\ell^{(2)}
=
a_1\widehat Y_\ell^{(2)}(\tau,r_{\ell2},r_{\ell3}),
\end{equation}
and denoting the singular values of $\widehat Y_\ell^{(2)}$ by
$\widehat s_1<\widehat s_2$, one has
\begin{equation}
|a_1|=\frac{y_\mu}{\widehat s_2},
\qquad
y_e^{\rm pred}=y_\mu\frac{\widehat s_1}{\widehat s_2}.
\end{equation}
The electron-to-muon hierarchy is therefore a genuine constraint on the
modular shape rather than an independent normalization.

\section{ Definitions of modular forms}\label{app:modforms}
\textcolor{black}{The modular forms used in the current analysis are defined as follows}
\begin{align}
    Y_1^{(4)} = Y_1^2 +Y_2^2 =S_Y,\quad Y_{2}^{(4)} = \begin{pmatrix}
        Y_2^2 - Y_1^2\\
        2Y_1Y_2
    \end{pmatrix} = \begin{pmatrix}
        D_Y\\
        P_Y
    \end{pmatrix}
\end{align}
\begin{align}
    &Y_1^{(6)} = Y_1(3Y_2^2 - Y_1^2),\quad Y_{1'}^{(6)} = Y_2( 3Y_1^2 - Y_2^2),\quad Y_2^{(6)} =(Y_1^2 + Y_2^2) \begin{pmatrix}
        Y_1\\
        Y_2
    \end{pmatrix}=S_Y \begin{pmatrix}
        Y_1\\
        Y_2
    \end{pmatrix},\notag \\
    &Y_1^{(8)} = (Y_1^2 + Y_2^2)^2 =S_Y^2,\quad Y_2^{(8)} = (Y_1^2 + Y_2^2)\begin{pmatrix}
        Y_2^2 - Y_1^2\\
        2Y_1Y_2
    \end{pmatrix} = S_Y \begin{pmatrix}
        D_Y\\
        P_Y
    \end{pmatrix}~.
\end{align}

\section{Inverse-seesaw block diagonalization}
\label{app:iss}

Write the full neutral matrix as
\begin{equation}
\mathcal M_\nu=
\begin{pmatrix}
0&B\\
B^T&H
\end{pmatrix},
\qquad
B=(M_D\;\;0),
\end{equation}
with
\begin{equation}
H=
\begin{pmatrix}
0&M_R^T\\
M_R&\mu_S
\end{pmatrix}.
\end{equation}
Integrating out the heavy block gives the Schur complement
\begin{equation}
m_\nu=-BH^{-1}B^T.
\end{equation}
To leading order in $\mu_S$,
\begin{equation}
H^{-1}=
\begin{pmatrix}
-M_R^{-1}\mu_SM_R^{-T} & M_R^{-1}\\
M_R^{-T} & 0
\end{pmatrix}
+\mathcal O(\mu_S^2),
\end{equation}
and therefore
\begin{align}
m_\nu
&=-(M_D\;\;0)
\begin{pmatrix}
-M_R^{-1}\mu_SM_R^{-T} & M_R^{-1}\\
M_R^{-T} & 0
\end{pmatrix}
\begin{pmatrix}M_D^T\\0\end{pmatrix}
\nonumber\\
&=
M_DM_R^{-1}\mu_SM_R^{-T}M_D^T.
\end{align}
Because $M_D$ has two columns,
\begin{equation}
{\rm rank}(m_\nu)\le {\rm rank}(M_D)\le2,
\end{equation}
which establishes the massless-neutrino prediction at leading order.

The same expansion gives the active-heavy rotation
\begin{equation}
F=M_DM_R^{-1},
\end{equation}
and the charged-current mixing matrix is approximately
\begin{equation}
N\simeq
\left(1-\frac12FF^\dagger\right)U_\nu,
\end{equation}
so $\eta_{\rm NU}\simeq FF^\dagger/2$.

In the $\mu_S\to0$ limit,
\begin{equation}
H_0=
\begin{pmatrix}
0&M_R^T\\
M_R&0
\end{pmatrix},
\end{equation}
and the four heavy Majorana eigenstates combine into two degenerate Dirac
pairs with masses equal to the singular values of $M_R$.  A nonzero
$\mu_S$ lifts each degeneracy and produces the spectrum shown in
Fig.~\ref{fig:heavy}.

\section{Pseudo-Dirac dark eigenvalues and ultraviolet relations}
\label{app:dark}

The characteristic equation of Eq.~\eqref{eq:Mchi} is
\begin{equation}
\lambda^2-\mu_\chi\lambda-M_\chi^2=0,
\end{equation}
with exact eigenvalues
\begin{equation}
\lambda_\pm=
\frac12
\left[
\mu_\chi\pm\sqrt{\mu_\chi^2+4M_\chi^2}
\right].
\end{equation}
For $|\mu_\chi/M_\chi|\ll1$,
\begin{equation}
\sqrt{\mu_\chi^2+4M_\chi^2}
=
2M_\chi\left[
1+\frac{\mu_\chi^2}{8M_\chi^2}
+\mathcal O\!\left(\frac{\mu_\chi^4}{M_\chi^4}\right)
\right].
\end{equation}
After rephasing the negative eigenvalue,
\begin{equation}
m_{\chi_1}\simeq|M_\chi|-\frac{|\mu_\chi|}{2},
\qquad
m_{\chi_2}\simeq|M_\chi|+\frac{|\mu_\chi|}{2},
\end{equation}
so $\delta\simeq|\mu_\chi|$. 

With
\begin{equation}
A=|y_DY_{\mathbf1}^{(4)}(\tau)|,
\qquad
B=|y_MY_{\mathbf1}^{(6)}(\tau)|,
\end{equation}
the mass relations are
\begin{equation}
m_\chi=A\frac{v_\Phi^2}{\Lambda},
\qquad
\delta=B\frac{v_\Phi^3}{\Lambda^2}.
\end{equation}
Eliminating $\Lambda$,
\begin{equation}
\Lambda=A\frac{v_\Phi^2}{m_\chi},
\end{equation}
and hence
\begin{equation}
\delta=
\frac{B}{A^2}\frac{m_\chi^2}{v_\Phi}.
\end{equation}
The exact ultraviolet relations are therefore
\begin{equation}
v_\Phi=
\frac{B}{A^2}\frac{m_\chi^2}{\delta},
\qquad
\Lambda=
\frac{B^2}{A^3}\frac{m_\chi^3}{\delta^2}.
\end{equation}

\section{Endothermic xenon kinematics and recoil shape}
\label{app:kin}

For
$\chi_1(p)+N(k)\rightarrow\chi_2(p')+N(k')$, define
$q=k'-k$ and
\begin{equation}
E_R=\frac{q^2}{2m_N}.
\end{equation}
Nonrelativistic energy conservation gives
\begin{equation}
\bm v\cdot\bm q=
\frac{q^2}{2\mu_{\chi N}}+\delta.
\end{equation}
For fixed $q$ the minimum incoming speed occurs for
$\bm v\parallel\bm q$, yielding
\begin{equation}
v_{\min}
=
\frac{1}{q}
\left(
\frac{q^2}{2\mu_{\chi N}}+\delta
\right).
\end{equation}
Using $q=\sqrt{2m_NE_R}$ reproduces Eq.~\eqref{eq:vmin}.

Minimizing $v_{\min}(E_R)$ with respect to $E_R$ gives the absolute
opening condition
\begin{equation}
\delta\le\frac12\mu_{\chi N}v_{\max}^2.
\end{equation}
Solving $v_{\min}(E_R)=v_{\max}$ for the two roots gives
Eq.~\eqref{eq:ERpm}.

For a contact interaction,
\begin{equation}
\frac{dR}{dE_R}
=
N_T\frac{\rho_\chi}{m_\chi}
\int_{v>v_{\min}}d^3v\,
vf(\bm v)\frac{d\sigma}{dE_R}.
\end{equation}
Factoring out the total normalization leaves
\begin{equation}
\frac{dR}{dE_R}\propto
F_{\rm Helm}^2(E_R)\eta(v_{\min}),
\end{equation}
where
\begin{equation}
\eta(v_{\min})
=
\int_{|\bm v|>v_{\min}}\frac{f(\bm v)}{v}\,d^3v.
\end{equation}
For a truncated Maxwellian halo,
\begin{equation}
z=\frac{v_{\rm esc}}{v_0},
\qquad
N_{\rm esc}={\rm erf}(z)-\frac{2z}{\sqrt\pi}e^{-z^2}.
\end{equation}
For $v_{\min}<v_{\rm esc}-v_E$,
\begin{align}
\eta(v_{\min})=
\frac{1}{2N_{\rm esc}v_E}
\Bigg[
&{\rm erf}\!\left(\frac{v_{\min}+v_E}{v_0}\right)
-{\rm erf}\!\left(\frac{v_{\min}-v_E}{v_0}\right)
\nonumber\\
&-\frac{4v_E}{\sqrt\pi v_0}e^{-z^2}
\Bigg].
\end{align}
For
$v_{\rm esc}-v_E\le v_{\min}<v_{\rm esc}+v_E$,
\begin{align}
\eta(v_{\min})=
\frac{1}{2N_{\rm esc}v_E}
\Bigg[
&{\rm erf}(z)
-{\rm erf}\!\left(\frac{v_{\min}-v_E}{v_0}\right)
\nonumber\\
&-\frac{2(v_{\rm esc}+v_E-v_{\min})}{\sqrt\pi v_0}e^{-z^2}
\Bigg],
\end{align}
and $\eta=0$ above $v_{\rm esc}+v_E$.

The Helm form factor is
\begin{equation}
F_{\rm Helm}(q)=
3\frac{j_1(qR_1)}{qR_1}e^{-(qs)^2/2},
\end{equation}
where a standard choice is
\begin{equation}
R=1.2A^{1/3}\ {\rm fm},
\qquad
s=0.9\ {\rm fm},
\qquad
R_1=\sqrt{R^2-5s^2}.
\end{equation}
The detector-level shape is obtained by the Gaussian convolution in
Eq.~\eqref{eq:smear}.

\section{Profiled two-bin likelihood}
\label{app:likelihood}

Let
\begin{equation}
S_i(m_\chi,\delta)=
\int_{B_i}dE_R\,R_{\rm sm}(E_R;m_\chi,\delta)
\end{equation}
be the predicted shape integral in bin $i$.  With
\begin{equation}
\lambda_i=\mu S_i,
\qquad
\mu\ge0,
\end{equation}
the Poisson log-likelihood is
\begin{equation}
\ln\mathcal L(\mu)=
\sum_i
\left[
n_i\ln(\mu S_i)-\mu S_i-\ln(n_i!)
\right].
\end{equation}
Define
\begin{equation}
N=\sum_i n_i,\qquad
S_{\rm tot}=\sum_iS_i.
\end{equation}
Then
\begin{equation}
\frac{\partial\ln\mathcal L}{\partial\mu}
=
\frac{N}{\mu}-S_{\rm tot},
\end{equation}
and the profiled estimator is
\begin{equation}
\widehat\mu=\frac{N}{S_{\rm tot}}
\end{equation}
when $S_{\rm tot}>0$.  Substitution gives
\begin{align}
\ln\mathcal L_{\rm prof}
={}&
\sum_i n_i\ln S_i
+
N\ln\left(\frac{N}{S_{\rm tot}}\right)
-N
-\sum_i\ln(n_i!)
\nonumber\\
={}&
\sum_i n_i\ln p_i+\text{constant},
\qquad
p_i=\frac{S_i}{S_{\rm tot}}.
\end{align}
For $(n_1,n_2)=(0,1)$,
\begin{equation}
\ln\mathcal L_{\rm prof}=\ln p_2+\text{constant}.
\end{equation}
After the total normalization is profiled, the simplified likelihood is
therefore driven by the fraction of the smeared spectral shape entering
the observed high-recoil bin.  This explains why the preferred region is
pushed toward the endothermic endpoint.

\section{Numerical scan envelopes and \texorpdfstring{positive-${\rm Re}\,\tau$}{positive-Re tau} branch}
\label{app:scan}

A true two-dimensional profile in the modular plane would require
\begin{equation}
\chi^2_{\rm prof}(\tau)=
\min_{\{r_{\ell2},r_{\ell3},r_{c2},r_{c3},r_d,r_{e2}\}}
\chi^2_{\ell+\nu},
\end{equation}
which is a 12-real-dimensional minimization at every fixed value of
$\tau$.  The production figure instead uses the $181447$ actually
accepted V9 solutions to construct conservative scan envelopes.  Accepted
points are assigned to the nearest of the ten refined basins; their
$\chi^2$ values are interpolated only inside the supported cloud, and
cells outside the convex hull or farther than the numerical-support mask
are excluded.  Thus the filled areas are scan-supported envelopes rather
than a fixed-$\tau$ confidence construction.

The scan explicitly includes dedicated positive- and
negative-${\rm Re}\,\tau$ global searches and refines both signs locally.
Consequently, the positive-real branches are part of the same dense
accepted sample rather than being shown only as isolated global minima.
\bibliographystyle{apsrev4-2}
\bibliography{references}

@misc{LZ2026HighRecoil,
  author = {{LUX-ZEPLIN Collaboration}},
  title = {Search for dark matter p@misc interactions in an extended nuclear recoil energy window with the {LUX-ZEPLIN} ({LZ}) experiment},
  eprint = {2609.02823},
  archivePrefix = {arXiv},
  primaryClass = {hep-ex},
  year = {2026}
}

@misc{SuYangYang2026,
  author = {Su, Liangliang and Yang, Jin Min and Yang, Wen-Na},
  title = {Inelastic Dark Matter Signature at High Recoil Energy in {LUX-ZEPLIN} and {CRESST}},
  eprint = {2609.01475},
  archivePrefix = {arXiv},
  primaryClass = {hep-ph},
  year = {2026}
}

@misc{DiMauro2026,
  author = {Di Mauro, Mattia},
  title = {Dark Matter at the Kinematic Edge: Interpreting the 248 keV {LZ} Nuclear-Recoil Candidate},
  eprint = {2609.02608},
  archivePrefix = {arXiv},
  primaryClass = {hep-ph},
  year = {2026}
}

@misc{Yamashita2026,
  author = {Yamashita, Kimiko},
  title = {Inelastic Dark Photon Dark Matter for the {LUX-ZEPLIN} High-Recoil Event and the Galactic Halo Gamma-Ray Excess},
  eprint = {2609.02868},
  archivePrefix = {arXiv},
  primaryClass = {hep-ph},
  year = {2026}
}

@misc{SmirnovGriffithBeacom2026,
  author = {Smirnov, Juri and Griffith, Spencer and Beacom, John F.},
  title = {Inelastic Signatures of Electroweak Dark Matter},
  eprint = {2609.04144},
  archivePrefix = {arXiv},
  primaryClass = {hep-ph},
  year = {2026}
}

@misc{Unwin2026,
  author = {Unwin, James},
  title = {Axion Portal Dark Matter and the {LUX-ZEPLIN} High-Recoil Event},
  eprint = {2609.04186},
  archivePrefix = {arXiv},
  primaryClass = {hep-ph},
  year = {2026}
}

@misc{BaerBarger2026,
  author = {Baer, Howard and Barger, Vernon},
  title = {Exothermic dark matter and the 248 keV nuclear recoil in {LUX-ZEPLIN}},
  eprint = {2609.06153},
  archivePrefix = {arXiv},
  primaryClass = {hep-ph},
  year = {2026}
}

@misc{LeeBaryonFlavor2026,
  author = {Lee, Hyun Min},
  title = {Inelastic dark matter and baryon flavor symmetry in light of {LUX-ZEPLIN} ({LZ}) experiment},
  eprint = {2609.06171},
  archivePrefix = {arXiv},
  primaryClass = {hep-ph},
  year = {2026}
}

@misc{DasKarmakarMahapatraPaul2026,
  author = {Das, Pritam and Karmakar, Biswajit and Mahapatra, Satyabrata and Paul, Partha Kumar},
  title = {Inelastic Self-interacting Dark Matter and {LUX-ZEPLIN} 248 keV Event in a {Dirac} Modular Inverse Seesaw},
  eprint = {2609.06825},
  archivePrefix = {arXiv},
  primaryClass = {hep-ph},
  year = {2026}
}

@misc{AhmedLeontaris2026,
  author = {Ahmed, Waqas and Leontaris, George K.},
  title = {A Dark-Dimension Origin of Geometric Inelastic Dark Matter: The {LUX-ZEPLIN} High-Recoil Event and Multi-Target Tests},
  eprint = {2609.07138},
  archivePrefix = {arXiv},
  primaryClass = {hep-ph},
  year = {2026}
}

@misc{BorahSingletDoublet2026,
  author = {Borah, Debasish and Sahoo, Sujit Kumar and Sahu, Narendra and Sharma, Shashwat},
  title = {Inelastic Singlet-Doublet Fermion Dark Matter in light of the 248 keV {LZ} event},
  eprint = {2609.07800},
  archivePrefix = {arXiv},
  primaryClass = {hep-ph},
  year = {2026}
}

@misc{LeeRandall2026,
  author = {Lee, Vincent S. H. and Randall, Lisa},
  title = {A Warped Extra Dimensional Candidate for the {LZ} 248 keV Event},
  eprint = {2609.09136},
  archivePrefix = {arXiv},
  primaryClass = {hep-ph},
  year = {2026}
}

@misc{LeeYoun2026,
  author = {Lee, Seung J. and Youn, Taewook},
  title = {Mixing-suppressed inelastic dark matter: a minimal model for the {LZ} 248 keV event},
  eprint = {2609.09138},
  archivePrefix = {arXiv},
  primaryClass = {hep-ph},
  year = {2026}
}

@misc{FanHeWangZhao2026,
  author = {Fan, Zi-Tong and He, Hong-Jian and Wang, Yu-Chen and Zhao, Yue},
  title = {Inelastic Dark Matter and High-Energy Recoil Signatures in {LZ}},
  eprint = {2609.10491},
  archivePrefix = {arXiv},
  primaryClass = {hep-ph},
  year = {2026}
}

@misc{NguyenLindenHooper2026,
  author = {Nguyen, Thong T. Q. and Linden, Tim and Hooper, Dan},
  title = {Solar Neutrino Constraints on Inelastic Dark Matter Scattering in Light of Recent {LUX-ZEPLIN} Observations},
  eprint = {2609.11833},
  archivePrefix = {arXiv},
  primaryClass = {hep-ph},
  year = {2026}
}

@misc{OkadaShigekamiWu2026,
  author = {Okada, Hiroshi and Shigekami, Yoshihiro and Wu, Jia-Jun},
  title = {Can a minimal radiative seesaw explain the {LZ} 248 keV event?},
  eprint = {2609.13038},
  archivePrefix = {arXiv},
  primaryClass = {hep-ph},
  year = {2026}
}

@misc{BorahRadiativeDirac2026,
  author = {Borah, Pankaj and Mahapatra, Satyabrata and Nath, Newton and Paul, Partha Kumar},
  title = {Inelastic Dark Matter at {LZ} from Radiative {Dirac} Neutrino Mass Paradigm},
  eprint = {2609.15027},
  archivePrefix = {arXiv},
  primaryClass = {hep-ph},
  year = {2026}
}

@misc{HeHongJeesun2026,
  author = {He, Xiao-Gang and Hong, Xuan and Jeesun, Sk},
  title = {Hadrophilic inelastic freeze-in dark matter in $q_1-q_2$ gauge extension and the high energy {LZ} event},
  eprint = {2609.15714},
  archivePrefix = {arXiv},
  primaryClass = {hep-ph},
  year = {2026}
}

@misc{PalmisanoTammaroTesi2026,
  author = {Palmisano, Stefano and Tammaro, Michele and Tesi, Andrea},
  title = {Inferring dark matter masses and interactions from high recoil energy events in {LUX-ZEPLIN}},
  eprint = {2609.15985},
  archivePrefix = {arXiv},
  primaryClass = {hep-ph},
  year = {2026}
}

@misc{SanninoTurner2026,
  author = {Sannino, Francesco and Turner, Jessica},
  title = {Interpreting the {LZ} 248 keV Event using Dark {QCD}},
  eprint = {2609.24988},
  archivePrefix = {arXiv},
  primaryClass = {hep-ph},
  year = {2026}
}

@misc{KhanEtAl2026,
  author = {Khan, Imtiaz and Capozziello, Salvatore and Mustafa, G. and Botirov, Farkhod and Abdujabbarov, Ahmadjon and Atamurotov, Farruh and Channuie, Phongpichit},
  title = {Elastic toroidal vector dark matter through a dark photon in the {LUX-ZEPLIN} high recoil window},
  eprint = {2609.25114},
  archivePrefix = {arXiv},
  primaryClass = {hep-ph},
  year = {2026}
}

@misc{DelepineKhalil2026,
  author = {Delepine, David and Khalil, Shaaban},
  title = {Model-Independent Sideband Constraints on Inelastic Dark Matter at the {LZ} High-Recoil Candidate},
  eprint = {2609.26698},
  archivePrefix = {arXiv},
  primaryClass = {hep-ph},
  year = {2026}
}

@misc{CentellesChuliaEtAl2026,
  author = {Centelles Chuliá, Salvador and Chattaraj, Ayan and Srivastava, Rahul and Srivastava, Surabhi},
  title = {Escaping Solar Capture: {Majoron} mediated Elastic Dark Matter for the {LZ} 248 keV Event},
  eprint = {2609.32435},
  archivePrefix = {arXiv},
  primaryClass = {hep-ph},
  year = {2026}
}

@misc{ChoudhuryDasSachdeva2026,
  author = {Choudhury, Debajyoti and Das, Jaydeb and Sachdeva, Divya},
  title = {Inelastic Dark Matter from a Higher-Dimensional Operator: The {LUX-ZEPLIN} High-Recoil Event and Thermal Relic},
  eprint = {2610.00645},
  archivePrefix = {arXiv},
  primaryClass = {hep-ph},
  year = {2026}
}

@misc{LouLu2026,
  author = {Lou, Yuanchao and Lu, Hsin-Ting},
  title = {Fermionic Dark Matter Absorption and the High-Energy Event in {LUX-ZEPLIN}},
  eprint = {2609.01592},
  archivePrefix = {arXiv},
  primaryClass = {hep-ph},
  year = {2026}
}

@misc{Feruglio2017,
  author = {Feruglio, Ferruccio},
  title = {Are neutrino masses modular forms?},
  eprint = {1706.08749},
  archivePrefix = {arXiv},
  primaryClass = {hep-ph},
  year = {2017}
}

@misc{PenedoPetcov2019,
  author = {Penedo, J. T. and Petcov, S. T.},
  title = {Lepton Masses and Mixing from Modular ${S_4}$ Symmetry},
  journal = {Nucl. Phys. B},
  volume = {939},
  pages = {292--307},
  year = {2019},
  eprint = {1806.11040},
  archivePrefix = {arXiv},
  primaryClass = {hep-ph}
}

@misc{NovichkovEtAl2019,
  author = {Novichkov, P. P. and Penedo, J. T. and Petcov, S. T. and Titov, A. V.},
  title = {Modular ${S_4}$ Models of Lepton Masses and Mixing},
  journal = {JHEP},
  volume = {04},
  number = {04},
  pages = {005},
  year = {2019},
  eprint = {1811.04933},
  archivePrefix = {arXiv},
  primaryClass = {hep-ph}
}

@misc{ZhangZhou2021,
  author = {Zhang, Xinyi and Zhou, Shun},
  title = {Inverse Seesaw Model with a Modular ${S_4}$ Symmetry: Lepton Flavor Mixing and Warm Dark Matter},
  journal = {JCAP},
  volume = {09},
  number = {09},
  pages = {043},
  year = {2021},
  eprint = {2106.03433},
  archivePrefix = {arXiv},
  primaryClass = {hep-ph}
}

@misc{MohapatraValle1986,
  author = {Mohapatra, R. N. and Valle, J. W. F.},
  title = {Neutrino Mass and Baryon-Number Nonconservation in Superstring Models},
  journal = {Phys. Rev. D},
  volume = {34},
  pages = {1642},
  year = {1986}
}

@misc{BernabeuEtAl1987,
  author = {Bernabeu, J. and Santamaria, A. and Vidal, J. and Mendez, A. and Valle, J. W. F.},
  title = {Lepton Flavor Nonconservation at High-Energies in a Superstring Inspired {Standard Model}},
  journal = {Phys. Lett. B},
  volume = {187},
  pages = {303--308},
  year = {1987}
}

@misc{MalinskyEtAl2009,
  author = {Malinsky, Michal and Ohlsson, Tommy and Xing, Zhi-zhong and Zhang, He},
  title = {Non-unitary neutrino mixing and CP violation in the minimal inverse seesaw model},
  journal = {Phys. Lett. B},
  volume = {679},
  pages = {242--248},
  year = {2009},
  eprint = {0905.2889},
  archivePrefix = {arXiv},
  primaryClass = {hep-ph}
}

@misc{AbadaArcadiLucente2014,
  author = {Abada, Asmaa and Arcadi, Giorgio and Lucente, Michele},
  title = {Dark Matter in the minimal Inverse Seesaw mechanism},
  journal = {JCAP},
  volume = {10},
  number = {10},
  pages = {001},
  year = {2014},
  eprint = {1406.6556},
  archivePrefix = {arXiv},
  primaryClass = {hep-ph}
}

@misc{BlennowEtAl2023,
  author = {Blennow, Mattias and Fernandez-Martinez, Enrique and Hernandez-Garcia, Josu and Lopez-Pavon, Jacobo and Marcano, Xabier and Naredo-Tuero, Daniel},
  title = {Bounds on lepton non-unitarity and heavy neutrino mixing},
  eprint = {2306.01040},
  archivePrefix = {arXiv},
  primaryClass = {hep-ph},
  year = {2023}
}

@misc{FernandezMartinezEtAl2016,
  author = {Fernandez-Martinez, Enrique and Hernandez-Garcia, Josu and Lopez-Pavon, Jacobo},
  title = {Global constraints on heavy neutrino mixing},
  journal = {JHEP},
  volume = {08},
  number = {08},
  pages = {033},
  year = {2016},
  eprint = {1605.08774},
  archivePrefix = {arXiv},
  primaryClass = {hep-ph}
}

@misc{NuFIT60,
  author = {Esteban, Ivan and Gonzalez-Garcia, M. C. and Maltoni, Michele and Martinez-Soler, Ivan and Pinheiro, Joao Paulo and Schwetz, Thomas},
  title = {{NuFit-6.0}: Updated global analysis of three-flavor neutrino oscillations},
  eprint = {2410.05380},
  archivePrefix = {arXiv},
  primaryClass = {hep-ph},
  year = {2024}
}

@misc{TuckerSmithWeiner2001,
  author = {Tucker-Smith, David and Weiner, Neal},
  title = {Inelastic dark matter},
  journal = {Phys. Rev. D},
  volume = {64},
  pages = {043502},
  year = {2001},
  eprint = {hep-ph/0101138},
  archivePrefix = {arXiv}
}

@misc{TuckerSmithWeiner2005,
  author = {Tucker-Smith, David and Weiner, Neal},
  title = {The Status of inelastic dark matter},
  journal = {Phys. Rev. D},
  volume = {72},
  pages = {063509},
  year = {2005},
  eprint = {hep-ph/0402065},
  archivePrefix = {arXiv}
}

@misc{BramanteEtAl2016,
  author = {Bramante, Joseph and Fox, Patrick J. and Kribs, Graham D. and Martin, Adam},
  title = {The Inelastic Frontier: Discovering Dark Matter at High Recoil Energy},
  journal = {Phys. Rev. D},
  volume = {94},
  pages = {115026},
  year = {2016},
  eprint = {1608.02662},
  archivePrefix = {arXiv},
  primaryClass = {hep-ph}
}

@misc{AnandFitzpatrickHaxton2014,
  author = {Anand, Nikhil and Fitzpatrick, A. Liam and Haxton, W. C.},
  title = {Model-independent Analyses of Dark-Matter P@misc Interactions},
  journal = {Phys. Rev. C},
  volume = {89},
  pages = {065501},
  year = {2014},
  eprint = {1308.6288},
  archivePrefix = {arXiv},
  primaryClass = {hep-ph}
}

@misc{Lewin:1995rx,
    author = "Lewin, J. D. and Smith, P. F.",
    title = "{Review of mathematics, numerical factors, and corrections for dark matter experiments based on elastic nuclear recoil}",
    reportNumber = "RAL-TR-95-024",
    journal = "Astropart. Phys.",
    volume = "6",
    pages = "87--112",
    year = "1996"
}

@misc{Holdom:1985ag,
    author = "Holdom, Bob",
    title = {Two U(1)'s and Epsilon Charge Shifts},
    reportNumber = "UTPT-85-30",
    journal = "Phys. Lett. B",
    volume = "166",
    pages = "196--198",
    year = "1986"
}

@misc{FabbrichesiGabrielliLanfranchi2020,
  author = {Fabbrichesi, Marco and Gabrielli, Emidio and Lanfranchi, Gaia},
  title = {The Dark Photon},
  journal = {SpringerBriefs in Physics},
  year = {2020},
  eprint = {2005.01515},
  archivePrefix = {arXiv},
  primaryClass = {hep-ph}
}

@article{Froggatt:1978nt,
    author = "Froggatt, C. D. and Nielsen, Holger Bech",
    title = "{Hierarchy of Quark Masses, Cabibbo Angles and CP Violation}",
    reportNumber = "CERN-TH-2519",
    doi = "10.1016/0550-3213(79)90316-X",
    journal = "Nucl. Phys. B",
    volume = "147",
    pages = "277--298",
    year = "1979"
}

@misc{Planck2018,
  author = {{Planck Collaboration}},
  title = {Planck 2018 results. VI. Cosmological parameters},
  journal = {Astron. Astrophys.},
  volume = {641},
  pages = {A6},
  year = {2020},
  eprint = {1807.06209},
  archivePrefix = {arXiv},
  primaryClass = {astro-ph.CO}
}

@misc{KhanAhmedLiRazaMuhammad2026,
  author        = {Khan, Imtiaz and Ahmed, Waqas and Li, Tianjun and
                   Raza, Shabbar and Muhammad, Ali},
  title         = {Thermal Higgsino Dark Matter in GmSUGRA for the LZ Event
                   and Solar-Neutrino Constraint},
  year          = {2026},
  eprint        = {2610.00226},
  archivePrefix = {arXiv},
  primaryClass  = {hep-ph},
    journal       = {arXiv preprint}
}

@misc{AhmedLeontarisKK2026,
  author        = {Ahmed, Waqas and Leontaris, George K.},
  title         = {Kaluza--Klein Dark-Photon Mediation of Inelastic Dark
                   Matter at LUX-ZEPLIN},
  year          = {2026},
  eprint        = {2609.22739},
  archivePrefix = {arXiv},
  primaryClass  = {hep-ph},
    journal       = {arXiv preprint}
}

@misc{AhmedAhmadRehman2026,
  author        = {Ahmed, Waqas and Ahmad, Ammara and Ur Rehman, Mansoor},
  title         = {Xenon Isotope Filtering at the Kinematic Edge of
                   Inelastic Dark Matter},
  year          = {2026},
  eprint        = {2609.15634},
  archivePrefix = {arXiv},
  primaryClass  = {hep-ph},
    journal       = {arXiv preprint}
}

@misc{McCabe2026,
  author        = {McCabe, Christopher},
  title         = {Seasonal dark matter from the LUX-ZEPLIN high-energy event},
  year          = {2026},
  eprint        = {2609.04181},
  archivePrefix = {arXiv},
  primaryClass  = {hep-ph},
  journal       = {arXiv preprint}

}

@misc{FreeseTheodosopoulos2026,
  author        = {Freese, Katherine and Theodosopoulos, Dionysios P.},
  title         = {Higgsino dark matter compatible with the LUX-ZEPLIN high-energy nuclear-recoil event and IceCube constraints},
  year          = {2026},
  eprint        = {2610.03692},
  archivePrefix = {arXiv},
  primaryClass  = {hep-ph}
}

@misc{Langhoff2026,
  author        = {Langhoff, Kevin},
  title         = {Heavy Higgsino Interpretation of the LZ Event},
  year          = {2026},
  eprint        = {2609.09385},
  archivePrefix = {arXiv},
  primaryClass  = {hep-ph}
}

@misc{PospelovRamani2026,
  author        = {Pospelov, Maxim and Ramani, Harikrishnan},
  title         = {Strong Constraints on Higgsino Dark Matter from Solar Capture},
  year          = {2026},
  eprint        = {2609.02775},
  archivePrefix = {arXiv},
  primaryClass  = {hep-ph}
}

@misc{BoseEtAlHiggsino2026,
  author        = {Bose, Debajit and Saha, Akash Kumar and Das, Deep Jyoti and Sherpa, Rinchen and Singh, Abhijeet and Ghosh, Durba and Doliya, Jaya and Bouri, Subhadip and Mondal, Biprajit and Mondol, Ranjini and Raj, Nirmal and Laha, Ranjan},
  title         = {Not so good $\nu$s for Higgsino dark matter as LZ excess: stringent limits from Super-Kamiokande and IceCube},
  year          = {2026},
  eprint        = {2609.07807},
  archivePrefix = {arXiv},
  primaryClass  = {hep-ph}
}

@misc{Pallis2026HeavyHiggsino,
  author        = {Pallis, C.},
  title         = {Heavy Higgsino Dark Matter With Low Reheating in View of
                   the LZ High-Recoil Event},
  year          = {2026},
  eprint        = {2609.33869},
  archivePrefix = {arXiv},
  primaryClass  = {hep-ph}
}

@misc{Wang:2020dbp,
    author = "Wang, Xin",
    title = "{Dirac neutrino mass models with a modular $S_4$ symmetry}",
    eprint = "2007.05913",
    archivePrefix = "arXiv",
    primaryClass = "hep-ph",
    journal = "Nucl. Phys. B",
    volume = "962",
    pages = "115247",
    year = "2021"
}

@misc{Basiouris:2024qfe,
    author = "Basiouris, Vasileios and Crispim Rom{\~a}o, Miguel and King, Stephen F. and Leontaris, George K.",
    title = "{Modular family symmetry in fluxed GUTs}",
    eprint = "2407.06618",
    archivePrefix = "arXiv",
    primaryClass = "hep-ph",
    reportNumber = "IPPP/24/42",
    doi = "10.1103/PhysRevD.111.015012",
    journal = "Phys. Rev. D",
    volume = "111",
    number = "1",
    pages = "015012",
    year = "2025"
}

@misc{Okada:2026fef,
    author = "Okada, Nobuchika and Raut, Digesh",
    title = "{Endothermic Z'-Portal Dark Matter: LZ-LHC Complementarity}",
    eprint = "2609.21011",
    archivePrefix = "arXiv",
    primaryClass = "hep-ph",
    month = "9",
    year = "2026"
}

\end{document}